%% file: Main.tex
\documentclass[aps,prl,twocolumn]{revtex4}
\usepackage{array}
\usepackage{graphicx}
\usepackage{epstopdf}
\usepackage{cancel}
\usepackage{comment}
\usepackage{natbib}
\usepackage[utf8]{inputenc}
\usepackage[dvipsnames]{xcolor}\usepackage{bm}
\usepackage{amsmath}
\usepackage[colorlinks=true,linkcolor=blue,urlcolor=blue,citecolor=blue]{hyperref}
\usepackage{tikz}
\usetikzlibrary{spy, patterns,patterns.meta, calc, backgrounds} 
\usepackage[outline]{contour}
\usepackage[position=top,margin=0pt,labelformat=parens,justification=justified]{subcaption}
\usepackage[normalem]{ulem}
\usepackage{caption}

\tikzset{>=latex} 
\usetikzlibrary{intersections}
\usetikzlibrary{calc}
\usetikzlibrary{decorations.markings}
\usetikzlibrary{angles,quotes} 
\colorlet{xcol}{black}
\tikzstyle{rvec}=[->,thick,xcol,line cap=round]

\usepackage[export]{adjustbox} 

\newcolumntype{P}[1]{>{\centering\arraybackslash}p{#1}}

\begin{document}
\title{Active Growth Layer Induced by Micromechanical Feedback \\ Shapes Proliferating Cell Collectives}

\author{Fidel \'Alvarez-Murphy${}^{1}$, Ignacio Medina${}^{1}$, N\'estor Sep\'ulveda${}^{2}$ and  Gustavo D\"uring${}^{1}$  }
\affiliation{${}^1$Instituto de F\'isica, Pontificia Universidad Cat\'olica de Chile, Casilla 306, Santiago, Chile }
\affiliation{${}^2$School of Engineering and Sciences,  Universidad Adolfo Ib{\'a}{\~n}ez, Diagonal las Torres 2640, Pe\~{n}alolen, Santiago, Chile.}

\begin{abstract}

Proliferating cell collectives often develop an active growth layer near their boundary that regulates expansion and morphology, as observed in systems ranging from bacterial biofilms to epithelial tissues and tumor spheroids. While such layers have been attributed to diverse mechanisms, their microscopic origin remains unclear in many situations. Here, we show that micromechanical feedback alone provides a minimal mechanism for their emergence. We introduce a particle-based model of non-motile proliferating cells in which growth is locally inhibited by compressive stress, coupling division to mechanical interactions and generating an active growth layer without biochemical regulation. An emergent mechanical length scale, denoted by $\chi$, sets the extent of the proliferative region and controls the system’s behavior across scales, governing growth dynamics, morphology and organizing internal stress and velocity fields.

Coarse-graining the model yields a continuum description with no adjustable parameters, providing a microscopic foundation for existing approaches. When the colony expands into a passive environment, we observe and characterize fingering instabilities driven purely by mechanical feedback. These instabilities can be tuned through the system geometry relative to $\chi$, and leads to an exponential acceleration of colony growth, enhancing the collective growth rate. We further establish a correspondence with nutrient-depletion models, providing a route to study the statistical properties of expanding fronts within a minimal microscopic framework.
\end{abstract}

\maketitle

\section{Introduction}
    

A hallmark of living systems is proliferation: living structures are composed of cells that grow, divide, and eventually die, giving rise to diverse and rich dynamical behaviors~\cite{Hallatschek2023}. These dynamics lead to the formation of evolving multicellular structures, including biofilms in the gut microbiome and tissues that develop into organs, or tumors in cases of uncontrolled growth~\cite{Rajendran2017,deVos2015,HallStoodley2004}. In nature, cell colonies often exhibit self-organizing behaviors driven by a variety of environmental cues such as nutrient availability, temperature, pH, and mechanical pressure~\cite{BenJacob1997,Goller2008}. These factors regulate proliferation and prevent unbounded expansion, ultimately giving rise to complex biological organization.

A common feature observed in experiments on growing cell collectives is the emergence of two distinct proliferative regions: an inner region with low division rates and elevated self-generated mechanical stress, and a peripheral active layer where cells continuously divide~\cite{Heinrich2020,Balmages2023,Trepat2009,Young2023,Su2012,Gauquelin2019,Asp2022}. The interaction between this proliferative front and the surrounding environment gives rise to a variety of morphological phenomena, ranging from finger-like instabilities~\cite{Lin2016,BenJacob1997}, to the formation of concentric ring patterns~\cite{Liu2021,BenJacob1997,Shimada2004}. Understanding and controlling the physical properties of this active front is essential in applications ranging from tissue morphogenesis~\cite{Friedl2009}, to the engineering of synthetic biological communities~\cite{Naseri2020,Barbier2022}, and strategies for controlling tumor invasion~\cite{Delarue2014,Montel2011}.
To describe these phenomena, multiple models have been proposed, including both continuum and discrete frameworks~\cite{Dockery2001,Ye2024,FarrellHallattschek2013,Alert2020,Giverso2016}, incorporating diverse feedback mechanisms for growth inhibition, such as nutrient depletion, programmed cell death (apoptosis), and mechanical stress~\cite{Dockery2001,FarrellHallattschek2013,Basan2009,Williamson2018,Ye2024}. In this work, we investigate how mechanical feedback alone shapes the dynamics of proliferating non-motile cell colonies through a minimal micromechanical model that couples cellular growth to internal stresses, as observed in experiments where growth is inhibited by mechanical stress accumulation~\cite{BenMeriem2023,Welker2021,Delarue2016}. We show that this mechanism yields an emergent active-layer length scale $\chi$ that sets spatial extent over which growth remains unsaturated.
This purely mechanical origin of $\chi$ contrasts with that in models where growth is regulated by nutrient availability~\cite{Dockery2001,Giverso2016}. Strikingly, the active layer thickness $\chi$ governs multiple aspects of the collective dynamics independently of its microscopic origin~\cite{Young2023}. We show that it controls the formation of ring-like patterns, the transition from exponential to linear expansion, and the characteristic wavelength of fingering instabilities. Moreover, the recently proposed mechanism whereby fingering instabilities enhance population growth~\cite{Ye2024} is also found to be governed by the same length scale, leading to an exponential acceleration of colony growth.

Starting from a discrete, particle-based model, we elucidate the origin and role of the active growth layer at the microscopic level. To probe further the structure of the active front and the emergence of instabilities, we derive a coarse-grained continuum description. These continuum equations extend previous theoretical formulations of pattern formation in proliferating systems~\cite{Ye2024,Dockery2001,Williamson2018}, linking local mechanical interactions to large-scale colony dynamics. In addition, this allows for a direct comparison with models in which growth is regulated by nutrient availability~\cite{Giverso2016}. Finally, we assess the scope and limitations of this continuum approach and of the model.

\section{Cell collective growth model modulated by mechanical feedback} 

We study a two-dimensional system of proliferating, soft, disk-shaped cells. Each cell \( i \) is defined by its position \( \vec{r}_i \) and radius \( a_i \). In the absence of mechanical interactions, cells grow at a constant rate \( \nu \) and divide once their radius reaches a fixed threshold \( a \). Upon division, a mother cell is replaced by two daughter cells that preserve the total area (see Methods for details). Mechanical feedback is modeled via repulsive forces that depend on the overlap between neighboring cells. These interactions capture both elastic repulsion and contact inhibition of proliferation: overlapping cells generate local stresses that reduce their growth rate, effectively reproducing pressure-mediated growth suppression.
\begin{figure*}[t]
\centering
    \includegraphics[width=\textwidth]{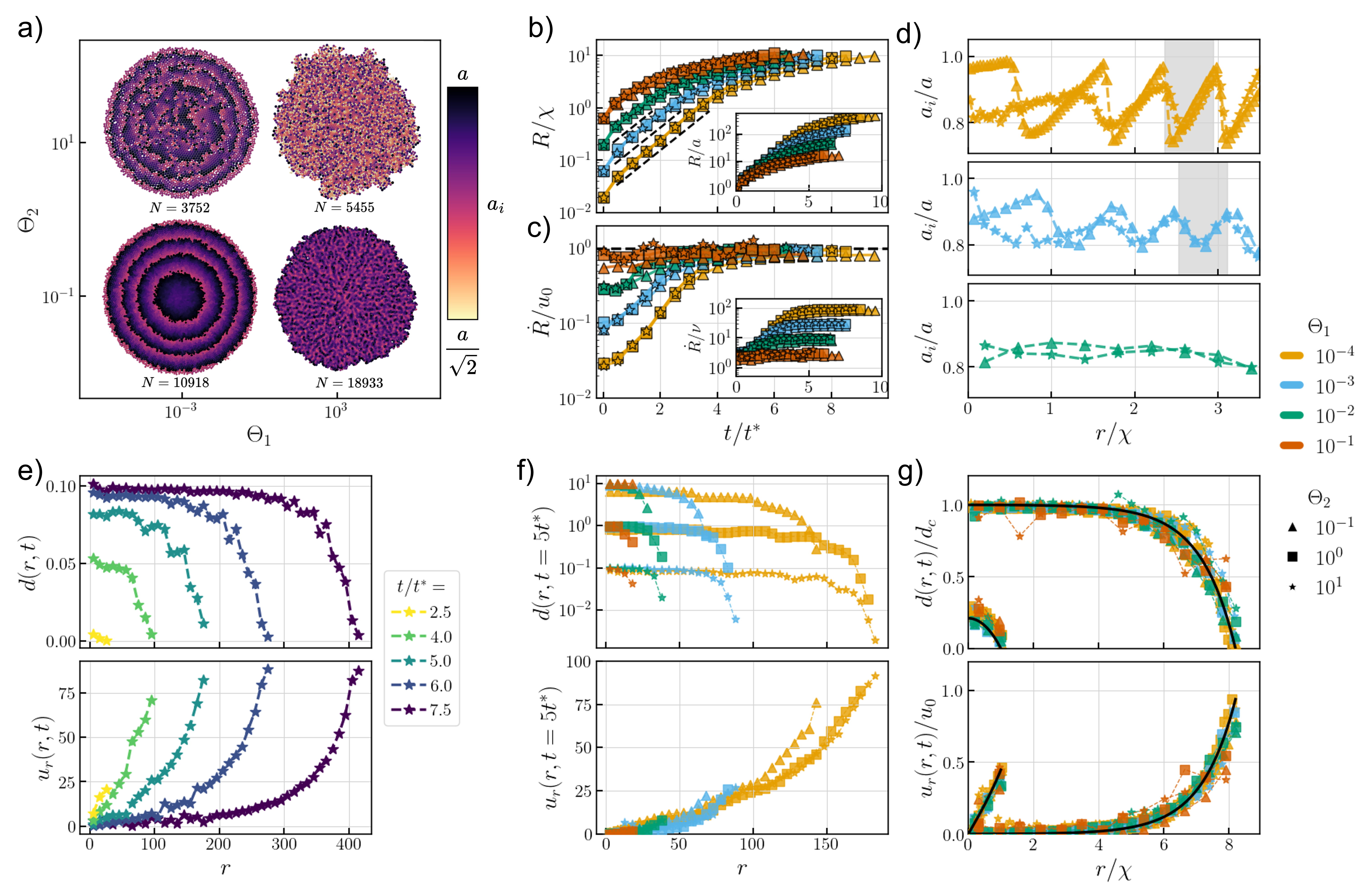}
\caption{Results from discrete simulations of proliferating colonies with different $\Theta_1$ and $\Theta_2$ in the free-boundary configuration. (a) Morphological phase diagram with a particle-size color map. (b,c) Rescaled growth $R(t)/\chi$ and front velocity $\dot{R}(t)/u_0$ curves  (dashed line in (b): $\sim e^{t/t^*}$; dashed line in (c): $\dot{R}/u_0=1$; insets: raw curves). (d) Radial distribution of cell sizes; shaded region indicates an interval of size $1-1/\sqrt{2}$. (e,f,g) Radial profiles of overlap and velocity: (e) temporal evolution for $\Theta_1=10^{-4}$, $\Theta_2=10$; (f) comparison of different parameter sets at $t=5t^*$; (g) collapse of radial profiles onto continuum theory predictions (black lines) at $R=\chi$ and $R=8\chi$.}
\label{fig:fig1}
\end{figure*}

The overlap between a pair of cells \( \{i,j\} \) is defined as \( d_{ij} = a_i + a_j - |\vec{r}_{ij}| \), with \( \vec{r}_{ij} = \vec{r}_j - \vec{r}_i \). Each cell has an effective dynamic viscosity \( \mu \) and an internal bulk viscosity \( \gamma \). The overdamped dynamics is given by
\begin{align}
\dot{\vec{r}}_i &= -\frac{K}{\mu a_i} \sum_{j \neq i} d_{ij} \, H\left( d_{ij} \right) \hat{r}_{ij}, \label{positionDynamics0} \\
\dot{a}_i &= \nu - \frac{K}{\gamma} \sum_{j \neq i} d_{ij} \, H\left(d_{ij} \right), \label{radiusDynamics0}
\end{align}
where \( H(\cdot) \) is the Heaviside step function, ensuring that only overlapping pairs ($d_{ij}>0$) contribute. Eq.~\eqref{positionDynamics0} describes displacement due to mechanical repulsion, while Eq.~\eqref{radiusDynamics0} captures the reduction of growth under compression.

Using \( a \) and \( \nu \) as characteristic length and velocity scales, we can define a natural time scale \( t^* = a/\nu \). Then a simple rescaling shows that the dynamical Eq.~\eqref{positionDynamics0} and Eq.~\eqref{radiusDynamics0} depend on two dimensionless parameters:
\[
\Theta_1 = \frac{\mu a}{\gamma} ,\qquad \Theta_2 = \frac{K a}{\gamma \nu}.
\]
We first simulate colony expansion in a free-boundary configuration for different values of the dimensionless parameters (Fig.~\ref{fig:fig1}a). Growth proceeds into an unconfined medium, forming a circular domain characterized by the colony radius \( R(t) \) and its front velocity \( \dot{R}(t) \). The unscaled data (insets in Figs.~\ref{fig:fig1}b,c) show an initial exponential growth regime set by the intrinsic growth rate $\nu$, followed by a crossover to constant front propagation at longer times.
Defining, for convenience, the characteristic length  $\chi \equiv a \Theta_1^{-1/2}/2$ and velocity  $u_0 \equiv \nu \Theta_1^{-1/2}$, we find in Figs.~\ref{fig:fig1}b,c that rescaling the radius by $\chi$ aligns the crossover between exponential and constant-velocity growth, while rescaling \( \dot{R}(t) \) by $u_0$ leads to a collapse of the front speed at long times. Thus, the crossover sets the emergent mechanical length scale of the system, while $u_0$ determines the terminal front velocity.

To better characterize the dynamics and morphology, we define two coarse-grained fields: the overlap $d(\vec{r},t)$ and the velocity field $\vec{u}(\vec{r},t)$, which represent, respectively, the average particle overlap and velocity in a neighborhood of $\vec{r}$ (see Methods Eqs. \eqref{eq:overlap-CG} and \eqref{eq:velocity-CG}).
Under a free-boundary configuration, the colony's radial symmetry implies that these fields depend only on the radial coordinate, $d(\vec{r},t)\equiv d(r,t)$ and $\vec{u}(\vec{r},t)\equiv u_r(r,t)\hat{r}$.

Fig.~\ref{fig:fig1}e shows the temporal evolution of overlap \( d(r,t) \) and radial velocity \( u_r(r,t) \) during colony growth. At early times, weak overlap leads to a velocity that increases from the center toward the front, reflecting outward motion driven by growth. As the colony expands over time, the bulk saturates at a critical overlap and becomes mechanically arrested ($u_r \approx 0$), while a boundary region with lower overlap sustains finite velocity. At a fixed time, Fig.~\ref{fig:fig1}f show how mechanical parameters controls the spatial profiles. $\Theta_1$ primarily sets the position and propagation of the advancing front, while $\Theta_2$ determines the bulk compression. Despite the different roles of $\Theta_1$ and $\Theta_2$, both the overlap and velocity fields exhibit a simple underlying structure. As shown in Fig.~\ref{fig:fig1}g for different colony sizes $R$, when profile distances are rescaled by $\chi$, overlaps by $d_c \equiv a\,\Theta_2^{-1}$ (which cancels growth equation~\eqref{radiusDynamics0}) and velocities by $u_0$, all profiles collapse onto universal curves. 
These universal profiles reveal a robust structure: the overlap decreases radially from a compressed bulk, where \( u \approx 0 \), toward an active outer layer of thickness \( \chi \), where low overlap enables cell proliferation and the velocity increases toward the front. A simple scaling argument then suggests that the front velocity scales with the number of growing cells per unit front surface times the single-cell growth rate, yielding \( u_0 \sim (\chi/a)\,\nu \), in agreement with the scaling observed in Figs.~\ref{fig:fig1}c. 

From Figs.~\ref{fig:fig1}a, one can also observe that variations in the mechanical parameters \( \Theta_1 \) and \( \Theta_2 \) lead to changes in the front roughness, as well as to the emergence of spatial ring-like patterns. The concentric ring-like structures observed in Fig.~\ref{fig:fig1}a, reminiscent of bacterial growth patterns~\cite{Shimada2004}, can be quantified through the radial dependence of the particle size $a_i$. These patterns arise from repeated cycles of compression and division, leading to a characteristic spacing set by the full growth-cycle time $t_c$, $u_0 t_c \sim \chi\left(1 - 1/\sqrt{2}\right)$, in agreement with the periodic modulation observed in Fig.~\ref{fig:fig1}d.
This structure is only present for $\Theta_1\ll 1$ ($\chi \gg a$). For larger $\Theta_1$ the radial modulation disappears and the colony loses this ordering. Within the patterned regime, $\Theta_2$ modifies the radial particle-size distribution near the center, reflecting changes in the internal stress distribution.

\begin{figure*}
\begin{center}
\includegraphics[width=0.75\textwidth]{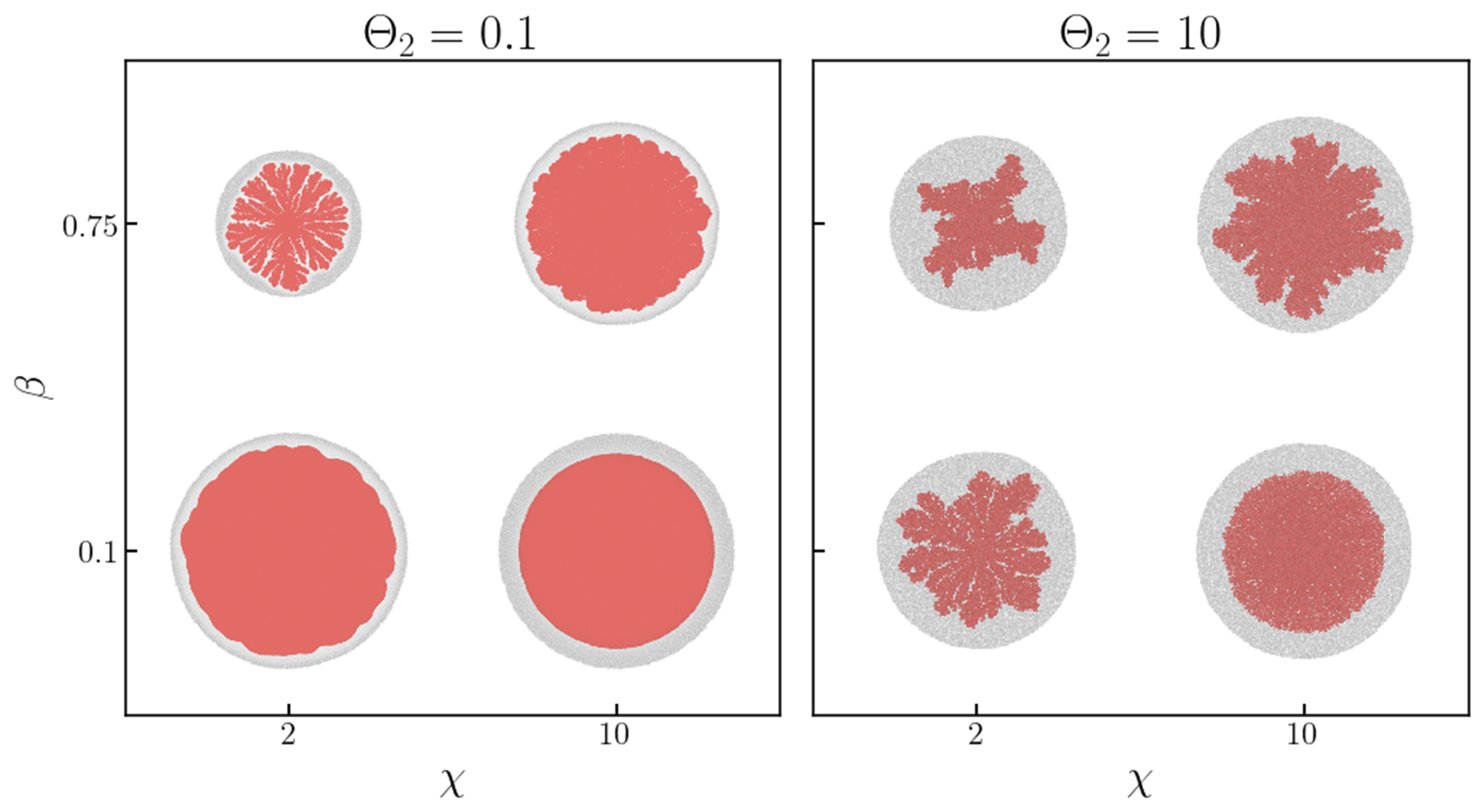}
\caption{Morphology of interfacial instabilities in a freely expanding colony as a function of $\chi$ and $\beta$, shown for $\Theta_2=\{0.1,10\}$.}
\label{fig:fig2}
\end{center}
\end{figure*}

While free expansion leads to stable front propagation across all parameter ranges, introducing a passive surrounding medium can trigger fingering instabilities (Fig.~\ref{fig:fig2}). This medium is modeled as a disordered collection of non-proliferative particles with radii \( a \) and \( a/\sqrt{2}\), sharing stiffness \(K\) with the active cells but with viscosity \(\mu_{\text{out}}\), and evolving according to Eq.~\eqref{positionDynamics0} with $\mu \to \mu_{\text{out}}$. The nature of the instability is governed by the viscosity ratio \( \beta = \mu_{\text{out}} / \mu \). 
For \( \beta > 1 \), classical fingering instabilities can emerge, reminiscent of the Saffman–Taylor mechanism~\cite{Saffman1958}. The case \( \beta < 1 \) is particularly interesting, as instabilities may emerge in a regime fundamentally different from the classical Saffman–Taylor mechanism. In this regime, pressure gradients induced by the outer medium stabilize the front, while proliferation drives the instability. The resulting morphology depends on \( \beta \), \( \Theta_1 \), and \( \Theta_2 \). Figure~\ref{fig:fig2} suggests that the characteristic length \( \chi \) governs both the onset and geometry of fingering. Smaller \( \chi \) produces narrower, faster-growing fingers, while larger \( \chi \) delays or suppresses the instability. Variations in $\Theta_2$ further modulate the shape of the protrusions.

Together, results in Figs.~\ref{fig:fig1}–\ref{fig:fig2} demonstrate that the same mechanical scale $\chi$, which controls colony growth, also sets the onset and geometry of interfacial instabilities. This unified behavior can be further explored within a continuum description.

\section{Coarse-Grained Theory, Front Dynamics, and Fingering Instability}

To analyze the formation and evolution of spatial structures in a proliferating colony, we develop a non-equilibrium continuum description based on the discrete dynamics defined in Eqs.~\eqref{positionDynamics0} and \eqref{radiusDynamics0}. Relying on a coarse-graining procedure we consider a representative mesoscopic region $d\Omega$ containing a sufficiently large number of particles to justify the local averaging of microscopic field quantities. In addition, since cell radius varies only within a narrow interval due to size conservation after division, from $\sim0.7a$ to $a$, all particles can be approximated as having a common size $a$ and a uniform reference density $\rho_0$. To make this resulting continuum description valid, it must be ensured that the overall colony in $\Omega$ contains many coarse-graining regions so particle-scale fluctuations are negligible at large scales. Tracking time-dependent mass variation in $d\Omega$, and incorporating the microscopic growth law into a coarse-grained continuity equation (see Supplementary Material), we arrive at the following evolution equation for the cell density field $\rho(\vec{r},t)$ in a non-motile colony:
\begin{equation}
\label{eq:ContinuityEq0}
\partial_t \rho(\vec{r},t) + \nabla\cdot\left[\rho(\vec{r},t) \vec{u}(\vec{r},t)\right] = \frac{2\nu}{a} \left[1 - \frac{d(\vec{r},t)}{d_c}\right] \rho(\vec{r},t),
\end{equation}
where $d(\vec{r},t)$ and $\vec{u}(\vec{r},t)$ denote the previously defined local overlap and velocity fields, respectively. Importantly, the characteristic critical overlap $d_c$ found in discrete observations (Fig.~\ref{fig:fig1}g), sets the threshold beyond which proliferation is fully suppressed in this formulation.

To link the velocity field $\vec{u}(\vec{r},t)$ to mechanical parameters, we consider the overdamped Cauchy momentum balance. Numerical measurements of virial stress tensor show that off-diagonal components remain negligible (see Supplementary Material), so the coarse-grained stress tensor is effectively isotropic. Under these conditions, momentum balance reduces to a Darcy-like law:
\begin{equation}
\label{eq:DarcyLaw}
\vec{u}(\vec{r},t) = -\zeta \nabla p(\vec{r},t),
\end{equation}
with mobility $\zeta = \pi a \rho_0 / (\mu \rho(\vec{r},t))$.

In the discrete model, mechanical stress is computed using the virial stress tensor. To relate pressure to overlap, we further assume that interparticle overlaps remain small compared to the cell diameter, corresponding to a small-overlap regime ($\Theta_2 \gg 1$). In this small-deformation limit, the virial stress is linearized in the overlap field, leading to the linear relation:
\begin{equation}
\label{eq:PresureOverlap}
p(\vec{r},t) = \frac{K \rho(\vec{r},t)}{2 a \pi \rho_0} d(\vec{r},t),
\end{equation}
yielding the critical pressure $p_c = \frac{K \rho(\vec{r},t)}{2a\pi\rho_0} d_c$.

\begin{figure*}
\includegraphics[width=1.8\columnwidth]{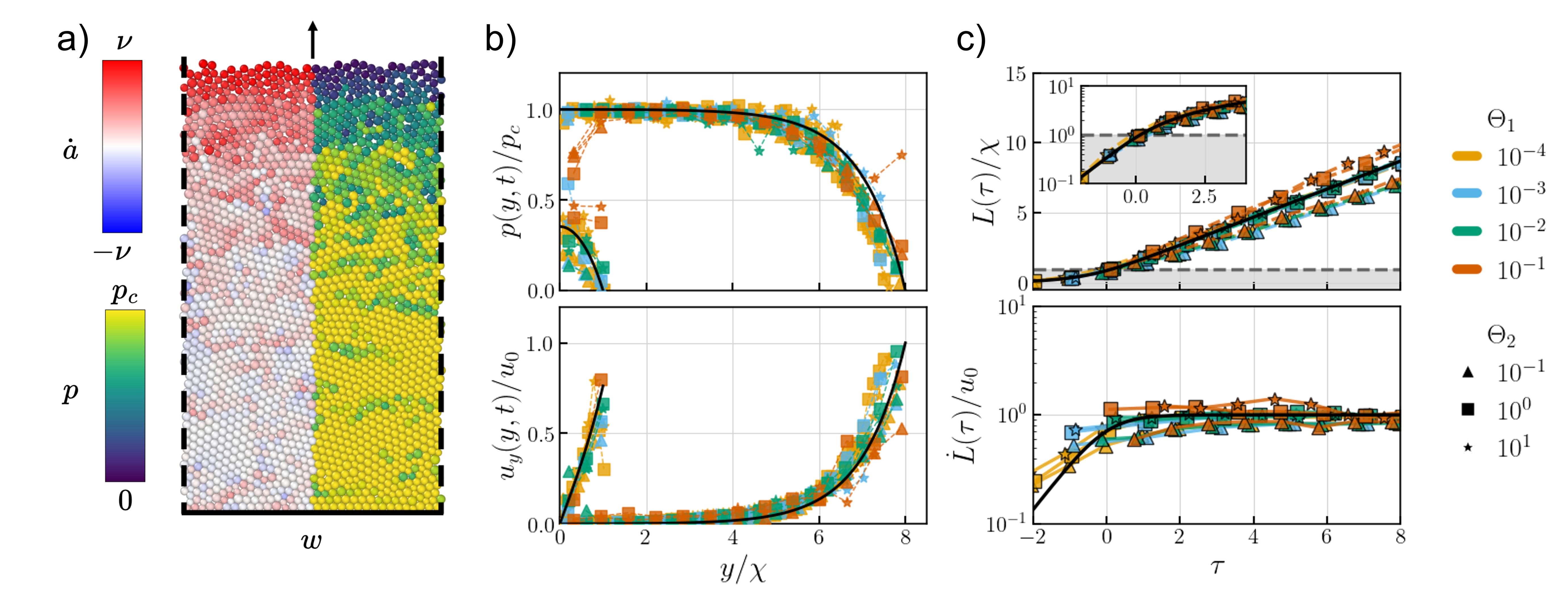}
\caption{Growth dynamics in a confined channel geometry. (a) Snapshot of the expanding colony in the channel-constrained configuration. (b) Collapse of longitudinal profiles onto continuum theory predictions (black lines) at $L=\chi$ and $L=8\chi$. (c) Rescaled growth $L(\tau)/\chi$ and front velocity $\dot{L}(\tau)/u_0$ curves compared to continuum theory predictions (black lines). Inset shows zoom in the exponential-linear growth transition}
\label{fig:fig3}
\end{figure*}

To close the system, we define a density profile using a level-set function $S(\vec{r},t)$, where $S<0$ represents the colony interior and $S = 0$ denotes the colony boundary. Because the particle size varies  weakly and is a nonmotile colony, density fluctuations inside the collective are small and the density can be approximated as constant $\rho_0$, therefore the density field is defined as:
\begin{equation}
\label{eq:GeneralDensityProf}
\rho(\vec{r},t) = \rho_0 \left(1 - H[S(\vec{r},t)]\right),
\end{equation}
where $H(\cdot)$ is the Heaviside step function. Substituting Eqs.~\eqref{eq:DarcyLaw}, \eqref{eq:PresureOverlap}, and \eqref{eq:GeneralDensityProf} into Eq.~\eqref{eq:ContinuityEq0} leads to an inhomogeneous Helmholtz equation for the pressure field:
\begin{equation}
\label{eq:InhHelmholtz}
\nabla^2 p(\vec{r},t) = \frac{1}{\chi^2} \left( p(\vec{r},t) - p_c \right).
\end{equation}
This equation shows that proliferation-induced stresses are screened over the characteristic length $\chi$, consistent with the simulations (Fig.~\ref{fig:fig1}b). The continuum formulation fixes the factor $2$ in the definition of $\chi$ introduced above (see Supplementary Material). Together, equations~\eqref{eq:DarcyLaw} and \eqref{eq:InhHelmholtz} provides a macroscopic description linking pressure, velocity and microscopic parameters, extending previous models such as~\cite{Ye2024}.

In the free-boundary geometry considered in Fig.~\ref{fig:fig1}, the colony expands radially with axisymmetric fields. Thus Eq. \eqref{eq:InhHelmholtz} can be solved in polar coordinates, determining the radial profiles of pressure and velocity that depend only on the instantaneous colony radius $R(t)$. The explicit derivation of these profiles is provided in the Supp. Material. Once the colony size exceeds $\chi$, the pressure saturates to $p_c$ in the bulk while the front propagates at a constant velocity $u_0$.
Using these analytical expressions, we compare the predicted profiles with those measured in the discrete simulations of the circular colony, where pressure is computed from the overlap via Eq.~\eqref{eq:PresureOverlap}. As shown in Fig.~\ref{fig:fig1}g, the continuum theory accurately reproduces the spatial structure of the mechanical fields for different colony sizes and parameters.

Now we have established this agreement, we turn to the dynamics of the colony boundary. The front position of the colony is determined by the kinematic boundary condition (KBC) obtained by integrating the continuity equation \eqref{eq:ContinuityEq0} across the moving boundary at $S=0$. This yields $\partial_t S+\vec{u} \cdot \nabla S =0$. For the circular expansion we found the front dynamics follows 
\begin{equation}
\label{eq:R(t)}
    \dot{R}(t)=u_0
\frac{I_1(R(t)/\chi)}{I_0(R(t)/\chi)},
\end{equation}
where $I_0$ and $I_1$ are modified Bessel functions of the first kind of order $0$ and $1$, respectively. This expression describes the observed exponential-to-linear crossover observed in Figs.~\ref{fig:fig1}b and~\ref{fig:fig1}c (See Supplementary Material for details).

A better description of the front dynamics can be achieved by considering colony growth along a straight channel of width $w$, with periodic boundary conditions in the transverse direction, as shown in Fig.~\ref{fig:fig3}a. In this configuration the colony advances along the channel, and its position is described by the front location $L(t)$. This setup provides a controlled framework to analyze both front dynamics and interfacial stability (see Supp. Material for details).
Not surprisingly, the resulting profiles for pressure and velocity retain the same screened structure identified in the free-boundary case, with a bulk saturation at pressure $p_c$ once $L \gg \chi$. As shown in Fig.~\ref{fig:fig3}b, the continuum predictions accurately reproduce the spatial structure of the mechanical fields across different colony sizes and parameters.
An explicit analytical solution for the front evolution can be obtained by combining the velocity field with the KBC, which yields
\begin{equation}
L(t) = \chi \, \mathrm{arcsinh} \left[ \exp \left(2 t/t^* \right) \sinh \left({L_0/\chi} \right) \right],
\label{eq:L(t)}
\end{equation}
where $L_0=L(t=0)$ is the initial colony length. This solution fully captures the crossover from exponential growth ($L \ll \chi$) to constant-velocity propagation $u_0$ ($L \gg \chi$). The predicted growth curves collapse under a suitable rescaling of time  $\tau = t / (2t^*) + \ln[\sinh(L_0/\chi)]$ and are in quantitative agreement with simulations without adjustable parameters, as shown in Fig.~\ref{fig:fig3}c for front position and velocity.

\begin{figure*}
\begin{center}
\includegraphics[width=\textwidth]{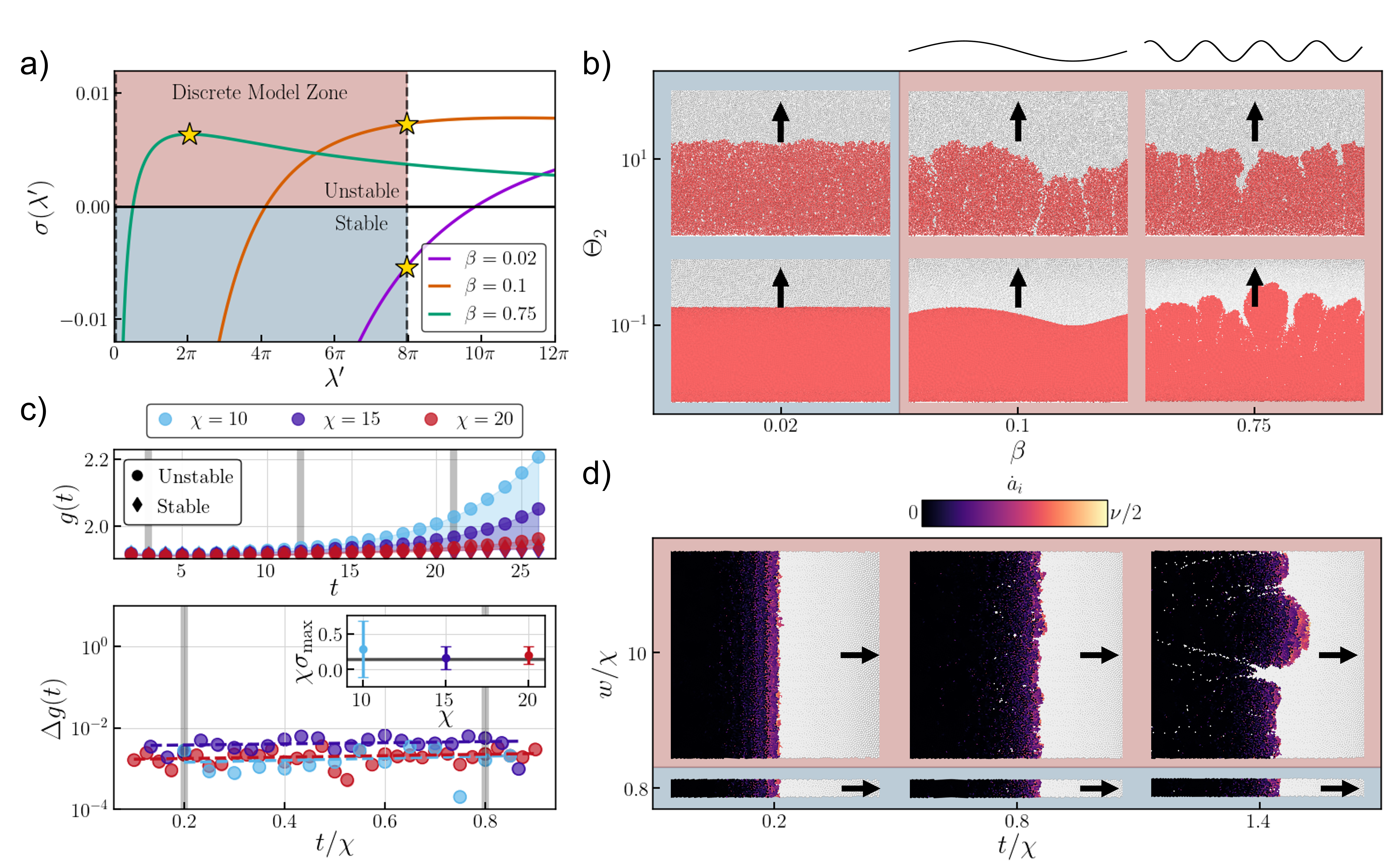}
\caption{(a) Dispersion relation \eqref{eq:DispersionRelation} for $\beta=\{0.02,0.1,0.75\}$ with $\chi=10a$ and $w=250a$. Shaded regions indicate the wavelengths accessible in the discrete model $\lambda' \in (\lambda'_{\min},\lambda'_{\max})$; stars mark the most unstable accessible modes $\sigma_\text{max}$. (b) Snapshots of colony fronts under the same conditions; superimposed sinusoidal profiles highlighting the dominant unstable wavelengths. (c) Top: Intensive growth rate $g(t)$ for stable $(w=12a)$ and unstable $(w\approx150a)$ configurations at $\chi=\{10a,15a,20a\}$ with $\beta=0.75$ and $\Lambda=5\chi$. Solid regions shows the difference $\Delta g(t)$. Bottom: Early-time $\Delta g(t)$ as function of rescaled time $t/\chi$. Dashed lines indicate exponential fits. Inset shows fitted growth exponent $\chi \sigma_\text{max}$ for each $\chi$ with error bars; solid line shows the theoretical prediction. (d) Time evolution of the colony front for unstable and stable configurations with $\chi=15a$, at the shaded times in (c).}
\label{fig:fig4}
\end{center}
\end{figure*}

In general, the continuum theory seems to agree well with simulations when a clear separation exists between the particle scale $a$ and the characteristic length $\chi$ (i.e., $\Theta_1 \ll 1$), where coarse-graining is well justified and the mechanical fields vary smoothly over many particles. As $\chi$ approaches the particle scale, deviations emerge due to the breakdown of scale separation, as observed for $\Theta_1 = 0.1$ (orange curves in Figs.~\ref{fig:fig1} and~\ref{fig:fig3}). In this regime the discrete effects become significant and the continuum description loses quantitative accuracy, although it still captures the overall scaling behavior. In addition, the channel configuration presents stronger deviations near the origin due to the presence of the immobile wall. Regarding the mechanical closure, although the continuum theory is formally derived in the limit $\Theta_2 \gg 1$, it remains accurate even down to $\Theta_2 = 1$. Minor discrepancies only appear for smaller values e.g., $\Theta_2 = 0.1$ (triangular symbols in Figs.~\ref{fig:fig1} and~\ref{fig:fig3}), where nonlinear contributions to the pressure become relevant and reduce the growth rate. In this regime the continuum theory slightly overestimates the front propagation speed, while still accurately capturing the spatial structure of the mechanical fields. 
Taken together, these results indicate that the continuum description is quantitatively accurate within the coarse-grained regime $\chi\gg a$, while remaining qualitatively robust beyond it in the explored parameters, even when nonlinearities or discrete effects become important.

%

We now investigate the dynamics of the expanding interface and the mechanical origin of finger-like instabilities during collective proliferation, a phenomenon observed in various biological systems~\cite{BenJacob1997,Lin2016}. We consider the interaction between a proliferating colony and a surrounding passive medium, modeled as a non-growing fluid occupying the region $S > 0$ (see Fig.~\ref{fig:fig2} and Fig.~\ref{fig:fig4}b,d), governed by the Laplace equation $\nabla^2 p(\vec{r},t) = 0$. %
We consider a channel geometry with interface $S=x-L(t)$ and neglect lateral confinement effects.
The passive medium is confined to a finite width $\Lambda \chi$, with $\chi$ the active length. 
Following Ref.~\cite{Ye2024}, we introduce a small perturbation to the interface of the form $\epsilon \xi(t) \exp(i k x)$, with wavelength $\lambda = 2\pi / k$ (see Supplementary Material for details). Defining $\sigma=\dot{\xi}/\xi$ and $\lambda' = \lambda / \chi$ the linear stability analysis yields the dispersion relation:
\begin{equation}
\label{eq:DispersionRelation}
    \sigma(\lambda') = \dfrac{1+(\beta-1)\sqrt{1+\dfrac{4\pi^2}{\lambda'^2}}}{2\pi+\beta\lambda'\sqrt{1+\dfrac{4\pi^2}{\lambda'^2}}\tanh\left(\dfrac{2\pi\Lambda}{\lambda'}\right)}  \dfrac{4\pi}{\Lambda\beta+1}  \dfrac{\nu}{a},
\end{equation}
where $\beta$ is the viscosity ratio defined earlier. This expression corresponds to a limiting case of Ref.~\cite{Ye2024}, obtained in the absence of surface tension and far from the channel boundaries. 
While Ref.~\cite{Ye2024} mainly focuses on the stability properties for $\beta \ge 1$ closer to the classical Saffman-Taylor viscous fingering scenario, our system naturally operates in the regime $\beta < 1$, where the colony is more viscous than the surrounding medium. In this case the instability is not driven by viscosity contrast alone but by growth-induced pressure gradients that destabilize the proliferating front, while the mechanical active length $\chi$ stabilizes short-wavelength perturbations.

Figure~\ref{fig:fig4}a shows the dispersion relation \eqref{eq:DispersionRelation} for representative $\beta < 1$ values. In this regime, the continuum theory generically predicts a band of unstable modes with a maximum growth rate $\sigma_\text{max}$ at a wavelength set by the screening length $\chi$. 
However, instabilities can only develop if these modes lie within the range accessible to the discrete system, bounded by $\lambda_{\min} \sim a$ and $\lambda_{\max} \sim w$. This constraint defines three regimes, as shown in Figs.~\ref{fig:fig4}a,b. If all accessible modes have negative growth rate, the front remains effectively stable, even though unstable modes exist in the continuum spectrum, as observed for $\beta=0.02$. If $\sigma_\text{max}$ occurs at the upper bound $\lambda'_{\max}$, the instability develops at the system size, leading to a single dominant finger as occurs for $\beta=0.1$. Finally, when $\sigma_\text{max}$ lies within $\lambda'_{\min} < \lambda' < \lambda'_{\max}$, the interface develops a characteristic fingering pattern with a wavelength selected by the dispersion relation, as in the case $\beta=0.75$. These scenarios quantitatively account for the morphologies observed in Fig.~\ref{fig:fig4}b, where the selected unstable wavelength sets the initial interfacial deformation, determining the number of emerging fingers.
While the dispersion relation is independent of $\Theta_2$, simulations reveal that for $\Theta_2 < 1$ increased compressibility prevents the expulsion of passive particles, whereas for $\Theta_2 > 1$ the stiffer response allows passive particles to infiltrate the active region. This highlights a nonlinear, $\Theta_2$-dependent coupling between phases beyond our linear continuum description.

Our results further show that front instabilities enhance colony growth by redistributing the local production rate along the interface. We define the intensive growth rate
\begin{equation}
    g(t)=\dfrac{1}{w\chi} \dfrac{dA}{dt},
\end{equation}
where $A$ is the total colony area and $w\chi$ defines an effective active region.
By varying the channel width $w$, we control the accessible wavelength range while keeping all mechanical parameters fixed, enabling a direct comparison between stable and unstable regimes.
As shown in Figs.~\ref{fig:fig4}c,d a narrow channel ($w=12a$) remains stable and shows a constant growth rate set by steady production within the active layer of thickness $\sim\chi$, whereas wider channels ($w\approx150a$), develops fingering instabilities and exhibit an exponential increase of $g(t)$.
To quantify this effect, we measure the difference between unstable and stable growth rates, $\Delta g\equiv g_\text{unstable}-g_\text{stable}$. At early times, we find
\begin{equation}
    \Delta g(t)\sim e^{2\sigma_\text{max}t},
\end{equation}
as shown in Fig.~\ref{fig:fig4}c. This scaling follows from a second-order expansion of the interface perturbation, where linear contributions vanish upon spatial averaging and the leading correction is quadratic in deformation amplitude $\xi$ (See Supplementary Material for details). Equivalently, this admits a simpler geometric interpretation. The effective proliferating region scales as $\sim\chi \mathcal{L}$ where $\mathcal{L}$ is the interfacial length. Then, small front undulations increase $\mathcal{L}$ as $\delta\mathcal{L}\sim\xi^2\sim e^{2\sigma_\text{max}t}$, yielding the observed growth enhancement.
Moreover, we find that $\chi\sigma_\text{max}$ remains approximately constant (see inset in Fig. ~\ref{fig:fig4}c), indicating that the active length scale $\chi$ controls not only the onset of the instability, but also its amplification rate and characteristic timescale over which it impacts global expansion.
This demonstrates that morphological instabilities, controlled here purely by geometry, can enhance collective proliferation under otherwise identical mechanical conditions. This mechanism complements previous proposals where instability-driven growth enhancement is controlled by interfacial tension~\cite{Ye2024}, showing that geometry alone can play a similar role.

Above stability analysis applies to confined geometries, where the accessible unstable modes are bounded by a fixed cutoff $\lambda_\text{max}\sim w$. By contrast, in freely expanding colonies the radial front position $R_I(t)$ sets the interfacial extent, yielding a time dependent cutoff $\lambda_\text{max}(t) \propto R_I(t)$ that increases monotonically.
Consequently, the accessible wavelength range continuously expands, eventually reaching the characteristic most unstable mode. 
Since the most unstable wavelength scales with $\chi$, characteristic instabilities emerge once this scale becomes accessible. Thus, $\chi$ sets the onset and characteristic size of interfacial patterns, as observed initially in Fig.~\ref{fig:fig2}.

\section{UNIFIED FRAMEWORK FOR FRONT PROPAGATION IN MECHANICALLY AND NUTRIENT-REGULATED SYSTEMS} 

\begin{figure}
\begin{center}
\includegraphics[width=\columnwidth]{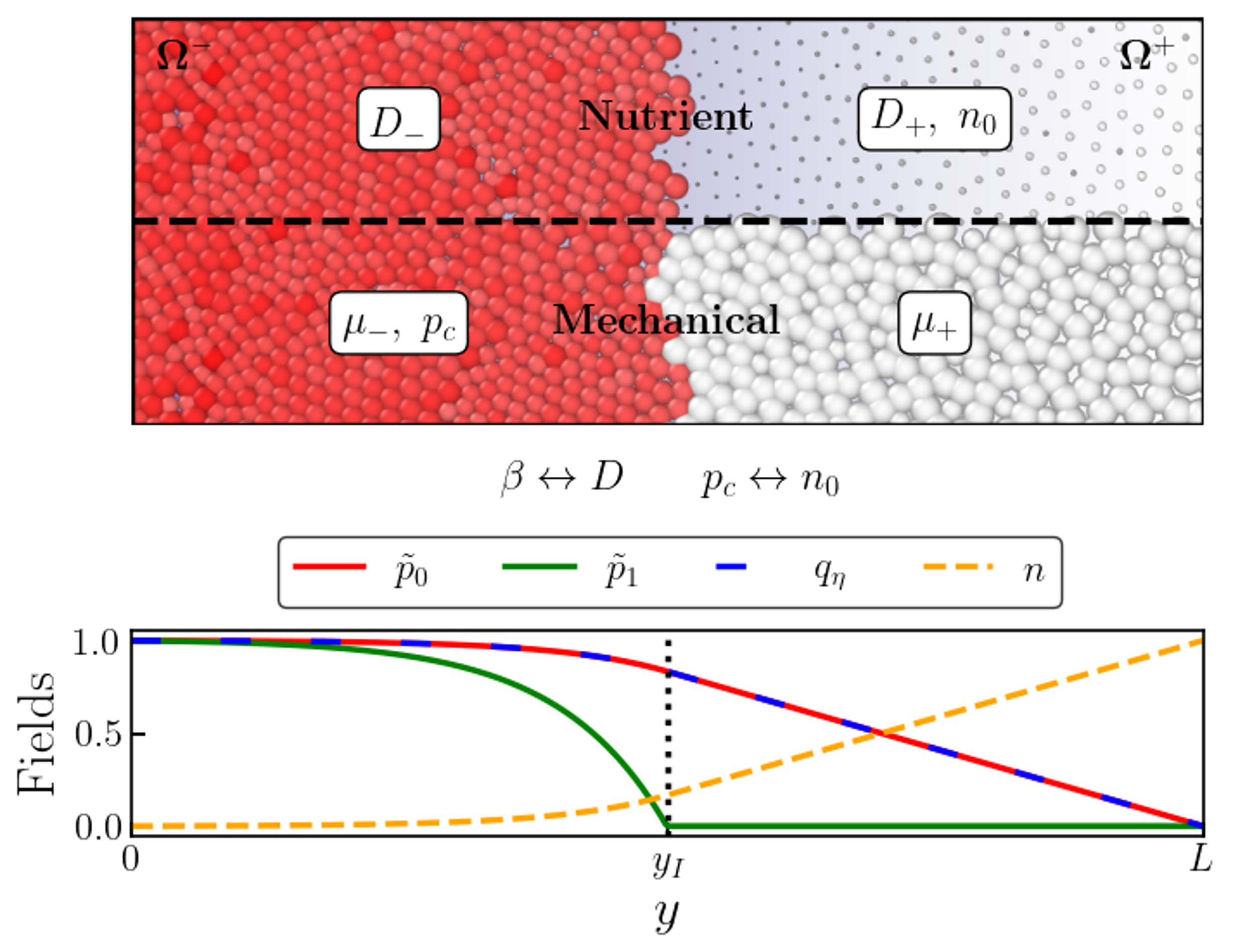}
\caption{Mapping between nutrient and mechanical descriptions of cell collectives in channel configuration. Top: illustration of both formulations in active  $(\Omega^-)$ and passive $(\Omega^+)$ media. Bottom: corresponding dimensionless field profiles $\tilde{p}_\eta$, $q_\eta$, and $n$ across the interface at $y_I$.}
\label{fig:fig5}
\end{center}
\end{figure}

A well-established class of biological growth models describes the expansion of cell colonies as a reaction-diffusion process regulated by nutrient consumption coupled to interfacial motion \cite{Dockery2001, Giverso2016}. In these models cells occupy an active region $\Omega^-$ while the surrounding domain $\Omega^+$ acts as a nutrient reservoir with external nutrient supply $n_0$. Controlling the whole dynamics, the nutrient field $n$ diffuses in $\Omega^+$ with diffusivity $D_+$ and penetrates into $\Omega^-$ with diffusivity $D_-$, where it is linearly consumed by proliferation and vanishing at the colony interior. This generates gradients through Darcy-like dynamics. At the continuum level, these models reduce to a scalar field satisfying Laplace dynamics in $\Omega^+$ and screened dynamics in $\Omega^-$, coupled by interfacial continuity conditions enforcing flux continuity with diffusivity contrast $D=D_+/D_-$.
Remarkably, our mechanically regulated model shares the same mathematical structure, despite a different physical interpretation of the surrounding medium, as illustrated in Fig.~\ref{fig:fig5}. In our case, $\Omega^-$ also defines the active region, characterized by a viscosity $\mu_-$, while the surrounding $\Omega^+$ is a passive material with viscosity $\mu_+$, transmitting mechanical stress rather than nutrients. The governing field is now the pressure, which extends across both regions, bounded by the saturation pressure $p_c$ and vanishing only at the outer boundary of the passive domain, obeying the same Laplace-screened dynamics as in the nutrient-driven case. Here, interfacial flux coupling is controlled by the viscosity contrast $\beta=\mu_+/\mu_-$, which plays the same structural role as the diffusivity ratio $D$ in the nutrient model.

This correspondence allows both descriptions to be cast into a unified formulation in terms of a scalar field $q_\eta$ driving interface motion and the spatial structure of growth, where $\eta\in\{0,1\}$ selects the physical mechanism. Therefore, $q_0\equiv p$ representing mechanical pressure, and $q_1 \equiv \alpha m$ nutrient deficit, with $m\equiv n_0-n$ and $\alpha$ a proportionality constant converting nutrient availability into an effective mechanical drive \cite{Giverso2016}. In this sense, we say mechanical pressure $p$ in our model acts as an effective realization of nutrient depletion $m$ in nutrient models with quasi-stationary evolution, governing growth through the same scalar variable. Therefore, the resulting field $q_\eta$ satisfies:
\begin{equation}
\nabla^2 q_\eta(\mathbf{r},t) =
\begin{cases}
0, & \mathbf{r}\in\Omega^+, \\[6pt]
\dfrac{1}{\chi^2}\,[q_\eta(\mathbf{r},t)-q_c], & \mathbf{r}\in\Omega^-,
\end{cases}
\label{eq:qfield}
\end{equation}
where $\chi$ is the screening length, determined by the internal flux parameters $\mu_-$ and $D_-$ in respective formulations, and $q_c$ sets the saturation source term in the active region. In the mechanical model $q_c=p_c$ corresponds to the saturating pressure, recovering \eqref{eq:InhHelmholtz} inside the colony, while in the nutrient-driven case $q_c = \alpha n_0$. The coupling between regions is encoded in the continuity of flux across the interface, $\kappa_\eta \nabla q_\eta(\partial \Omega^+) \cdot \hat{n}=\nabla q_\eta(\partial \Omega^-) \cdot \hat{n}$, where $\kappa_\eta$ is the contrast parameter. In nutrient-based models $\kappa_1 =D$, while in our mechanical model $\kappa_0 =\beta^{-1}$. Fig.~\ref{fig:fig5} illustrates the solutions of \eqref{eq:qfield} in a channel of length $L$, where passive medium extends up to $y=L$ and the active-passive interface is located at $y_I<L$, highlighting how mechanical based models pressure $p$ acts in the same way as the nutrient depletion $m$ from nutrient models. In contrast, in nutrient-driven systems the exterior $\Omega^{+}$ does not support mechanical stresses, and the effective pressure vanishes at the colony interface, while the diffusive field is defined in it. Thus, although both models can be described by a scalar field $q_\eta$, the physical realization of the pressure field differs in its spatial support.

The correspondence can be made explicitly by writing the effective pressure $\tilde{p}_\eta$ driving the interface in a unified form that links both mechanisms,
\begin{equation}
    \tilde{p}_\eta=q_\eta-\eta \left(
    \alpha n_0-\phi
    \right).
\end{equation}
Here, $\phi$ arises from expressing the nutrient-driven pressure $\tilde{p}_1$ field in terms of $n$, satisfying $\nabla^2 \tilde{p}_1=-\alpha \nabla^2 n$ for quasi-stationary evolution \cite{Giverso2016}. It corresponds to the solution $\nabla^2 \phi=0$ fixed by the boundary condition $\tilde{p}_1(y_I)=0$.
As illustrated in Fig.~\ref{fig:fig5}, our mechanical model pressure is recovered for $\tilde{p}_0=q_0=p$, while in nutrient formulation $\phi$ enforces the pressure $\tilde{p}_1$ to vanish at interface.
When $\phi$ is spatially uniform, as in this channel configuration, it does not contribute to pressure gradient, so mechanical and nutrient-driven models reduce to the same linear operator governing front propagation, becoming strictly linearly equivalent.
As a consequence, in the regime of stable or weakly perturbed fronts, both descriptions become dynamically indistinguishable, since the propagation speed, screening length $\chi$, and spatial structure of the velocity field are fully determined by the same linear theory. This provides a unified description of front propagation in mechanically and nutrient-regulated systems. In this sense, the mechanical model offers a minimal realization of the same effective theory, allowing to study statistical dynamics of steady-state front phenomena such as transport properties and gene surfing without explicitly introducing a nutrient field.

Differences arise beyond the leading-order dynamics. When curvature and higher-order perturbations are included, the $\eta$-coupled term contributes explicitly to the dispersion relation. As we did before, linearizing around a steady interface (see Supp for details), the general dispersion relation reads
\begin{equation}
    \sigma_\eta = -\zeta \left(
    \xi^{-1} \partial_y q_\eta^{(1)} + \partial_y^2 q_\eta^{(0)} - \dfrac{2\pi\eta}{\lambda}\left(
    \xi^{-1} q_\eta^{(1)} + \partial_y q_\eta^{(0)}
    \right)
    \right),
\end{equation}
showing that this additional coupling modifies how interface perturbations feed back into the front dynamics (See Fig. S6 from Supp material). Notice that choosing $\eta=0$ recovers Eq. \eqref{eq:DispersionRelation}.
As a result, for $\kappa_\eta>1$ ($\beta<1$) both models presents similar stability properties, with a finite band of unstable wavelengths set by $\chi$, although quantitative deviations arise from the $\eta$-coupled term. In contrast, for $\kappa_\eta<1$ nutrient-driven system keeps a finite unstable band, while the mechanical model becomes unstable down to arbitrarily small scales, with the fastest-growing mode shifting to $\lambda \to 0$ and hitting the particle size scale $\sim a$. However, as discussed earlier, finite-size effects such as confinement introduce a cutoff on accessible unstable modes. When this cutoff lies out the unstable band set by $\chi$, the nutrient-driven system is effectively stabilized. Under these conditions, both models recover equivalent stable propagating fronts, consistent with the linear correspondence discussed above. The differences become particularly evident in unstable configurations and beyond the linear regime, leading to distinct morphologies. Nutrient-based models can develop bubble-like protrusions associated with effective negative pressures, while in this mechanical model such collapse is not observed within the explored parameter range due to sustained particle growth.

\section{Conclusion}

Our results show that mechanical feedback alone is sufficient to generate rich spatial organization in continuously proliferating matter, including growth patterns, active boundary layers, and morphological instabilities, all governed by a single active length scale $\chi$ emerging from microscopic mechanical interactions. This active length, set by the ratio of elastic and viscous parameters, defines the range over which stress propagates and regulates proliferation, leading to a robust separation between a mechanically arrested bulk and an active expanding rim that drives expansion.

By coarse-graining the discrete cell dynamics, we derived a continuum description that quantitatively captures both the internal mechanical structure and the front dynamics directly from microscopic parameters. In both free-boundary and confined geometries, the theory predicts a crossover from exponential to constant-speed growth, as a consequence of the emergence of an expanding saturated bulk and a finite-size active layer which sets the expanding front velocity $u_0$. 
Beyond homogeneous expansion, this framework further explains that the same mechanical scale governs the onset of interfacial instabilities when the colony expands into a passive medium. Growth-induced pressure gradients destabilize the interface, while the finite screening length $\chi$ regularizes short-wavelength perturbations and selects a characteristic scale. Furthermore, when the system geometry allows unstable modes to develop, interfacial deformations redistribute growth toward protruding regions amplifying local activity and reinforcing front fluctuations, leading to fingering patterns that accelerate colony expansion. Consistently, the same scale also governs how rapidly these deformations translate into enhanced growth.
Finally, the equivalence between mechanically and nutrient-regulated models in the regime of stable fronts places our results within a broader class of interfacial growth phenomena. In this regime, both descriptions become dynamically indistinguishable, so that our mechanical framework provides a minimal model to study steady front dynamics without introducing additional fields, while directly linking macroscopic behaviors to microscopic parameters.


%

%

\section{Methods}

We simulate a minimal model of proliferating cells with mechanical feedback.
Each particle $i$ represents a cell with position $\vec{r}_i=(x_i,y_i)$ and radius $a_i$, and is characterized by mechanical parameters $\mu$, $\gamma$, and $K$, corresponding to substrate drag, internal dissipation, and elastic stiffness, respectively.

Visualization and rendering of simulation snapshots were performed using OVITO~\cite{ovito}.

\paragraph{\textbf{Simulation algorithm.}}
The system is evolved through an iterative time-stepping scheme:

\textit{(i) Division.}  At each iteration, every particle satisfying the division condition $a_i > a$ is removed and replaced by two daughters of radius $a/\sqrt{2}$, preserving area. The daughters are placed along a random axis at contact without overlapping.

\textit{(ii) Neighbor list.} A grid-based spatial partitioning is used to construct neighbor lists. After all divisions, particle neighbor lists are updated. Each particle considers potential interactions only with particles in its own and neighboring grid cells.

\textit{(iii) Overlaps.}  For each particle $i$, we compute the overlap with neighboring objects.

For a pair of particles $i$ and $j$, with relative position vector $\vec{r}_{ij}=\vec{r}_j-\vec{r}_i$, the overlap is defined as
\[d_{ij} = \min\!\left(0, |\vec{r}_{ij}| - (a_i + a_j)\right), 
\qquad 
\hat{n}_{ij} = \frac{\vec{r}_{ij}}{|\vec{r}_{ij}|},\]
where $\hat{n}_{ij}$ is the unit normal vector connecting particle centers.

For a horizontal wall at $y=0$, the overlap reads
\[
d_{i,w} = \min(0, y_i - a_i), 
\qquad 
\hat{n}_{i,w} = \hat{y}.
\]

\textit{(iv) Dynamics.} The instantaneous velocities of particle position and radius are computed as:
\[
\dot{\vec{r}}_i = \frac{K}{\mu a_i} \sum_j d_{ij} \hat{n}_{ij} + \frac{K}{\mu a_i} d_{i,w} \hat{n}_{i,w},
\]
\[
\dot{a}_i = \nu - \frac{K}{\gamma} \sum_j d_{ij} - \frac{K}{\gamma} d_{i,w},
\]
where the sum accounts for all particle-particle interactions.

\textit{(v) Position Update.} Particle positions and radii are updated using an explicit Euler scheme with timestep $\Delta t$
\[
\vec{r}_i(t + \Delta t) = \vec{r}_i(t) + \Delta t\, \dot{\vec{r}}_i, \quad a_i(t + \Delta t) = a_i(t) + \Delta t\, \dot{a}_i,
\]

\paragraph{\textbf{Colony size.}} For circular colonies, the effective radius is estimated from the spatial extent of the particles:
\[
R = \frac{1}{2}  \frac{\max(a, |x_i - x_j|) + \max(a, |y_i - y_j|)}{2},
\]
where the maxima are taken over all particle pairs ${i,j}$, and $a$ is the critical cell size. For channel geometries, the longitudinal extent is computed as:
\[
L = \max(y_i, a),
\]
with the maximum taken over all particles.

\paragraph{\textbf{Coarse-grained fields.}} 
Mechanical observables are obtained by spatial coarse-graining over a representative neighborhood $d\Omega$ centered at position $\vec{r}$ and containing $N_{d\Omega}$ cells. Within each region $d\Omega$, the coarse-grained overlap field is defined as
\begin{equation}
\label{eq:overlap-CG}
d(\vec{r},t) = \frac{1}{N_{d\Omega}} \sum_{i \in d\Omega} \sum_{j \neq i} d_{ij} \, H\left( d_{ij}  \right),
\end{equation}
where the sum runs over all particles $i$ inside $d\Omega$ and their interacting neighbors $j$, and the coarse-grained velocity field is computed as
\begin{equation}
\label{eq:velocity-CG}
\vec{u}(\vec{r},t) = \frac{1}{N_{d\Omega}} \sum_{i \in d\Omega} \dot{\vec{r}}_i,
\end{equation}
i.e., as the average particle velocity within the same region.

In practice, the system is partitioned into bins according to the geometry: concentric annuli for circular colonies and rectangular regions for channel geometries. The bin width is set to $5a$ for $\Theta_1<10^{-2}$ and to $a$ otherwise.

\paragraph{\textbf{Colony Area Measurement.}} 
We estimate the total colony area from the discrete representation as
\[
A(t) =  \sum_{i \in \text{Active}}\pi a_i^2,
\]
where the sum runs over all active particles.

\bibliographystyle{unsrtnat}

\bibliography{BiblioDef}

\clearpage
\onecolumngrid

\section{Supplementary Material}

\input{SuppAdd}

\end{document}

%% file: SuppAdd.tex
\setcounter{equation}{0}
\setcounter{figure}{0}
\setcounter{table}{0}

\renewcommand{\theequation}{S\arabic{equation}}
\renewcommand{\thefigure}{S\arabic{figure}}
\renewcommand{\thetable}{S\arabic{table}}

\section{S1. Discrete model}

\begin{figure}[h!]
\centering
\includegraphics[width=0.6\textwidth]{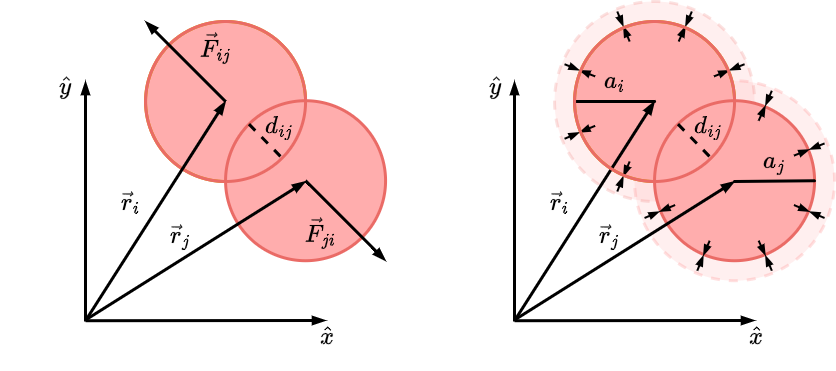}
\caption{Schematic representation of inter-cell interactions: mechanical repulsion (left) and growth's mechanical inhibition (right)}
\label{interactions}
\end{figure}

\noindent \textbf{Motion Equations.} — We consider a two-dimensional active cell described by its position $\vec{r}_i(t)$ and radius $a_i(t)$. The mechanical description includes three contributions: an active growth force $F_G$, viscous drag forces $F_D$, and inter-cell interaction forces $F_{ij}$ (see Fig.~\ref{interactions}). To account for both translation and growth, we introduce two equations of motion for $\vec{r}_i$ and $a_i$. Using Newton's second law, these read

\begin{align}
m\ddot{\vec{r}}_i & = \vec{F}_{D,r} + \vec{F}_{ij,r}, \label{position_dina}\\
m\ddot{a}_i &= F_G + F_{D,a} + F_{ij,a}. \label{radius_dina}
\end{align}

Here, $\vec{F}_{D,r}$ and $F_{D,a}$ denote the drag forces opposing translational motion and radial growth, respectively, while $\vec{F}_{ij,r}$ and $F_{ij,a}$ account for inter-cell interactions. 
We assume that growth occurs at a constant rate $\dot{a}_i = \nu$. For an isolated cell, the radial drag balances growth, with $F_{D,a} = -\gamma \dot{a}_i$, where $\gamma$ sets the stiffness scale. In the overdamped limit ($\ddot{a}_i = 0$) and in the absence of interactions ($F_{ij,a}=0$), Eq.~\eqref{radius_dina} yields a constant growth force

\begin{equation}
F_G = \gamma \nu.
\end{equation}

For the translational dynamics, cells are assumed non-motile and embedded in a viscous medium. The only contribution is a drag force opposing motion and proportional to the cell size,

\begin{equation}
\vec{F}_{D,r} = -\mu a_i \dot{\vec{r}}_i,
\end{equation}
where $\mu$ is an effective viscosity.

Cell-cell interactions arise upon overlap. Defining the overlap between cells $i$ and $j$ as

\begin{equation}
d_{ij} = a_i + a_j - r_{ij}, \label{eq:overlapdef}
\end{equation}
with $r_{ij} = |\vec{r}_j - \vec{r}_i|$, interactions occur when $d_{ij} > 0$. We model them as linear elastic forces $F_{ij} = -K d_{ij}$. The resulting interaction force on the center of mass is $\vec{F}_{ij,r} = F_{ij} \hat{r}_{ij}$, where $\hat{r}_{ij} = (\vec{r}_j - \vec{r}_i)/r_{ij}$, while the same magnitude contributes to radial dynamics, $F_{ij,a} = |\vec{F}_{ij,r}|$.

Substituting these expressions into Eqs.~\eqref{position_dina} and \eqref{radius_dina}, and taking the overdamped limit, we obtain Eqs. (1) and (2) from the main text:

\begin{align}
\vec{0} &= -\mu a_i \dot{\vec{r}}_i - K\sum_{j\neq i}^{N} d_{ij}H(d_{ij}), \label{positionDynamics0SUP}\\
0 &= - \gamma \dot{a}_i + \gamma \nu - K\sum_{j\neq i}^{N} d_{ij}H(d_{ij}), \label{radiusDynamics0SUP}
\end{align}
where $H(d_{ij})$ is the Heaviside function enforcing interactions only upon contact.

The system can be nondimensionalized by using the characteristic scales: length $r=a$, velocity $v=\nu$, and time $t=a/\nu$. This leads to the dimensionless parameters
\begin{align}
\Theta_1 &= \frac{\mu a}{\gamma}, \label{Theta1}\\
\Theta_2 &= \frac{K a}{\gamma \nu}. \label{Theta2}
\end{align}

In terms of these variables, Eqs.~\eqref{positionDynamics0SUP} and \eqref{radiusDynamics0SUP} become

\begin{align}
\dot{\vec{r}}_i &= -\frac{\Theta_2}{\Theta_1} \frac{1}{a_i} \sum_{j\neq i} d_{ij}H(d_{ij}) \hat{r}_{ij}, \label{positionDynamicsDimless2}\\
\dot{a}_i &= 1 - \Theta_2 \sum_{j\neq i} d_{ij}H(d_{ij}). \label{radiusDynamicsDimless2}
\end{align}

\begin{figure}[h!]
\centering
\includegraphics[width=0.8\textwidth]{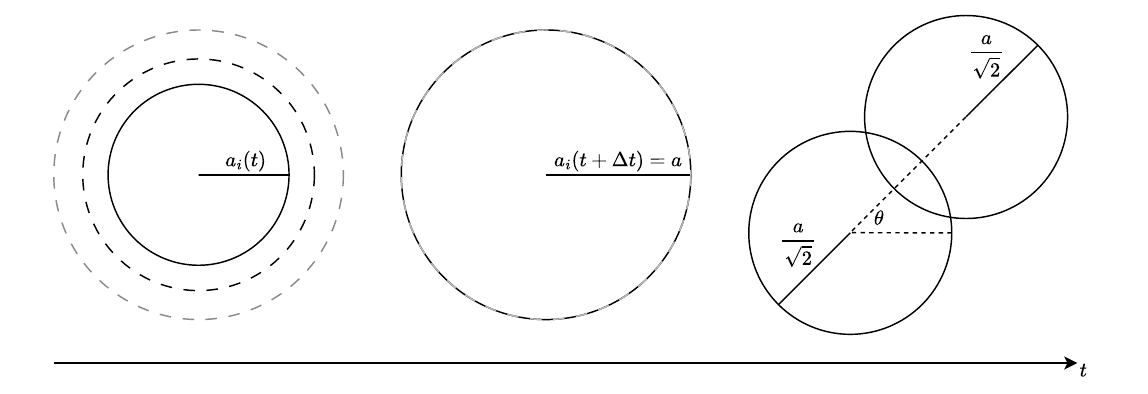}
\caption{Representative scheme of the cellular proliferation cycle}
\label{schematic_division_model}
\end{figure}

\noindent \textbf{ Division Algorithm.} — Cell proliferation is triggered when a cell reaches a critical size $a_i(t)=a$. We model division as binary fission, where one mother cell splits into two identical daughter cells (see Fig.~\ref{schematic_division_model}). 

Assuming constant density, mass conservation implies that the total area is preserved, yielding daughter, born at time $t_0$ with starting radii $a_j(t_0) = a/\sqrt{2}$. The division axis is defined by an angle $\theta \in [0,2\pi)$, chosen uniformly at random.

\section{S2. Formulation of continuous theory of growth colony}
\subsection{S2.1. Out-Of-Equilibrium Continuity Equation}

In this section, the continuum version of the colony's discrete model is derived.

\begin{figure}%
    \centering
    \includegraphics[width=6cm]{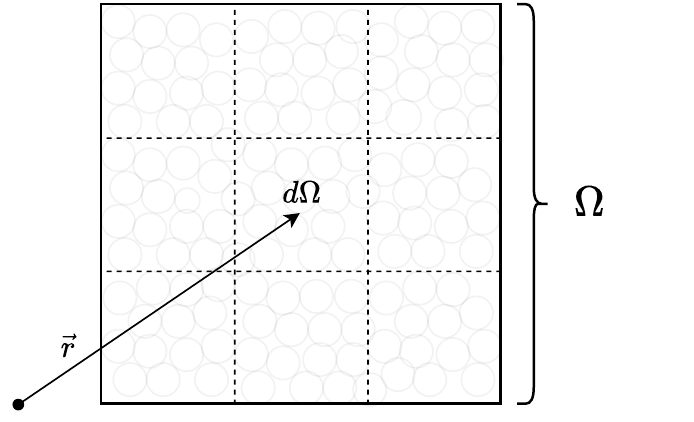}%
    \caption{Coarse grain of the region $\Omega$ divided in regions of area $d\Omega$.}%
\label{A2DColony}%
\end{figure}

A growing cellular colony is described by a density field $\rho(\vec r,t)$ and a velocity field $\vec u(\vec r,t)$. Consider a region $\Omega$ of area $\mathcal A$ containing many particles, such that a continuum description applies (Fig.~\ref{A2DColony}).

Mass conservation in $\Omega$ reads

\begin{equation}
\frac{dm}{dt} = \dot m_{\partial \Omega} + \frac{dm_{\text{gen}}}{dt},
\label{continuityeqsupp}
\end{equation}
where $\dot m_{\partial \Omega}$ is the net mass flux through the boundary and $dm_{\text{gen}}/dt$ accounts for mass production due to cell growth.

Building from the density field $\rho(\vec{r},t)$ the total mass rate is

\begin{equation}
\frac{dm}{dt} = \iint_{\Omega} \frac{\partial \rho(\vec r,t)}{\partial t}\, dS.
\label{massratedensity}
\end{equation}
where $dS$ is a differential element of surface

The mass flux across $\partial \Omega$, with $d\vec{l}$ is a differential element of line of the boundary $\partial \Omega$, is:

\begin{equation}
     \dot{m}_\Omega = \oint_{\partial\Omega}(\rho(\vec{r},t)\vec{u}(\vec{r},t))\cdot d\vec{l} = \iint_{\Omega}\nabla \cdot (\rho(\vec{r},t)\vec{u}(\vec{r},t))dS
\label{massfluxdivergence}
\end{equation}
where we used the divergence theorem.

\noindent \textbf{Mass generation.} — Cell growth increases particle areas. For small time interval $\Delta t$,

\begin{equation}\label{c7}
\Delta m_{gen} = \sum_{i\in\Omega} \rho(\vec{r}_i,t)[A_i(t+\Delta t) - A_i(t)],
\end{equation}
where $\rho(\vec{r}_i,t)$ is the density field around the position $\vec{r}_i$, $A_i$ is the area of the $i$-particle, and the sum is over all the particles inside the region $\Omega$. Since every cell has the same density $\rho_0$, thus the density field around the position $\vec{r}_i$ results $\rho(\vec{r}_i,t)=\rho_0$. Explicitly the area of the $i$-particle at time $t+\Delta t$ is $A_i(t+\Delta t) = \pi [a_i(t+\Delta t)]^2$. Therefore, keeping first-order terms on $\Delta t$,

\begin{equation}
\Delta m_{\text{gen}} = 2\pi \rho_0 \sum_{i\in\Omega} a_i \dot a_i\, \Delta t.
\label{c11}
\end{equation}
Defining the total overlap of the $i$-particle:

\begin{equation}
 d_i = \sum_{j\neq i} d_{ij}H\left( d_{ij} \right), 
 \label{overlapperparticle}
 \end{equation}
then Eq. \eqref{radiusDynamics0} leads to the growth law:

\begin{equation}
\dot{a}_i= \nu \left( 1 -\frac{d_i}{d_c} \right).
\label{growthlaw}
\end{equation}
where we've defined $d_c=\dfrac{K}{\gamma\nu}=a\Theta_2^{-1}$.
Replacing \eqref{growthlaw} in \eqref{c11}, we obtain

\begin{equation}
\Delta m_{\text{gen}} = 2\pi \rho_0 \nu \Delta t \sum_{i\in\Omega} a_i \left(1 - \frac{d_i}{d_c}\right).
\end{equation}
Assuming weak size fluctuations ($a_i \approx a$),\begin{equation}
\Delta m_{\text{gen}} = 2\pi \rho_0 a \nu \Delta t \sum_{i\in\Omega} \left(1 - \frac{d_i}{d_c}\right).
\end{equation}

\noindent \textbf{Coarse-graining.} — Dividing $\Omega$ with $N_{\Omega}$ particles into smaller regions $d\Omega_j$ of area $d\mathcal{A}_j$ with $N_{d\Omega_j}$ particles, we define the coarse-grained overlap field

\begin{equation}
d(\vec r,t) =  \dfrac{\sum_{i\in d\Omega_j} d_i}{N_{d\Omega_j}}.
\end{equation}
The generated mass term in the region $d\Omega_j$ results:

\begin{equation}\label{c13}
    \Delta m_{gen,d\Omega_j} 
    = 2\pi \rho_0 a \nu\Delta t N_{d\Omega_j}  \left[1 - \frac{d(\vec{r},t)}{d_c}\right]
\end{equation}
Since the particles in $\Omega$ have density $\rho_0$ and radius $\sim a$, we consider the density field as

\begin{equation}
    \rho(\vec{r},t) = \dfrac{\sum_{i\in d\Omega_j} \rho_0 \pi a^2 }{d\mathcal{A}_j} = \rho_0\pi a^2 \dfrac{N_{d\Omega_j}}{d\mathcal{A}_j}
\end{equation}
As long as $N_{\Omega}\gg N_{d\Omega}$, we can write:

\begin{equation}
\frac{dm_{\text{gen}}}{dt} = \frac{2\nu}{a} \iint_{\Omega} \left(1 - \frac{d(\vec r,t)}{d_c}\right)\rho(\vec r,t)\, dS.
\label{generatedmassflux}
\end{equation}

\noindent \textbf{Continuity equation.} — Replacing \eqref{massratedensity}, \eqref{massfluxdivergence} and  \eqref{generatedmassflux} into \eqref{continuityeqsupp}, leads to

\begin{equation}
    \iint_{\Omega}\frac{\partial \rho(\vec{r},t)}{\partial t} dS + \iint_{\Omega}\nabla\cdot(\rho(\vec{r},t) \vec{u}(\vec{r},t))dS = \frac{2\nu }{a}\iint_{\Omega} \left[1 - \frac{d(\vec{r},t)}{d_c}\right] \rho(\vec{r},t) dS
\end{equation}
Since this holds for arbitrary $\Omega$, we obtain

\begin{equation}\label{Acontinuity2}
\boxed{
    \frac{\partial \rho(\vec{r},t)}{\partial t} + \nabla\cdot[\rho(\vec{r},t) \vec u(\vec{r},t)] = \frac{2 \nu }{a} \left[1-\dfrac{d(\vec{r},t)}{d_c}\right]\rho(\vec{r},t)
    }
\end{equation}
which corresponds to continuity equation (5) in main text.

\subsection{S2.2. Constitutive relation between pressure and Velocity}

We now derive a constitutive relation linking the velocity field $\vec u(\vec r,t)$ to the mechanical pressure of the colony.

\begin{figure}[!h]
  \centering
\includegraphics[width=0.4\textwidth]{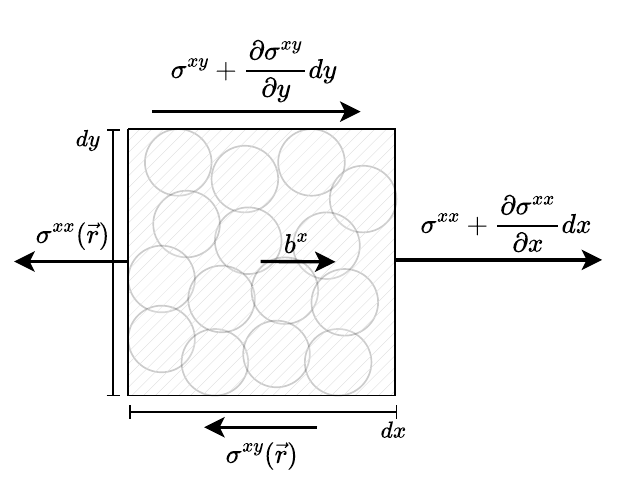}
\includegraphics[width=0.4\textwidth]{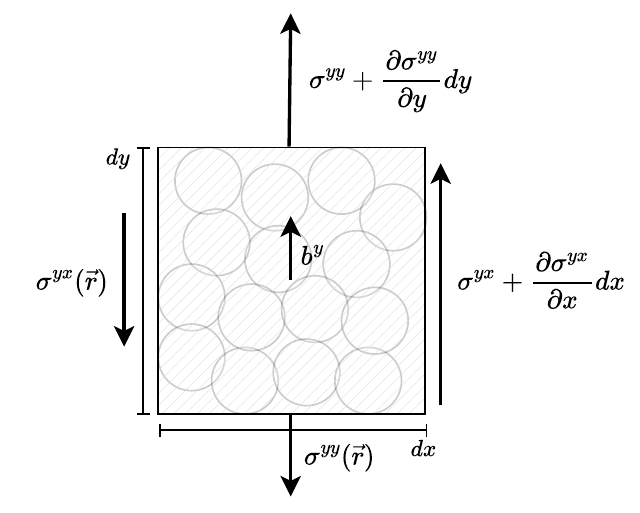}
  \caption{Two dimensional stresses and body forces acting on a control area $d \mathcal{A} = d x d y$, on the left stresses simplified to show effects on $\hat{x}$ and on the right for $\hat{y}$.}
  \label{A2Dstresses}
\end{figure}

Consider a coarse-grained element $d\Omega$ of area $d\mathcal A = dx\,dy$, subject to body forces $\vec b$ and surface forces encoded in the stress tensor $\pmb{\sigma}$ (Fig.~\ref{A2Dstresses}). In the overdamped regime, force balance reads

\begin{equation}
\partial_\beta \sigma^{\alpha\beta} + b^\alpha = 0,
\label{A2DNewton}
\end{equation}
where the sum over repeated indices is implied, $\partial_\alpha=\frac{\partial}{\partial \alpha}$ and $\alpha,\beta \in \{x,y\}$.

\noindent \textbf{Body force.} — At the particle level, dissipation is dominated by substrate drag, $\vec b_i = -\mu a \dot{\vec r}_i$. Coarse-graining over $d\Omega$ yields

\begin{equation}
\vec b(\vec r,t) = -\frac{\mu \rho(\vec r,t)}{\rho_0 \pi a}\,\vec u(\vec r,t),
\label{A2Dbodyforce1}
\end{equation}
where the velocity field is defined as the local average

\begin{equation}
    \vec{u}(\vec{r},t) = \frac{\sum_{i\in d\Omega} \dot{\vec{r}}_i}{N_{d\Omega}}
\end{equation}

\noindent \textbf{Stress tensor.} — In general, $\pmb{\sigma}$ contains both normal and shear components. The off-diagonal terms $\sigma^{xy}$ and $\sigma^{yx}$ encode shear stresses arising from local anisotropies in cell packing and interactions. 

\begin{figure}[h!]
\centering
\includegraphics[width=\textwidth]{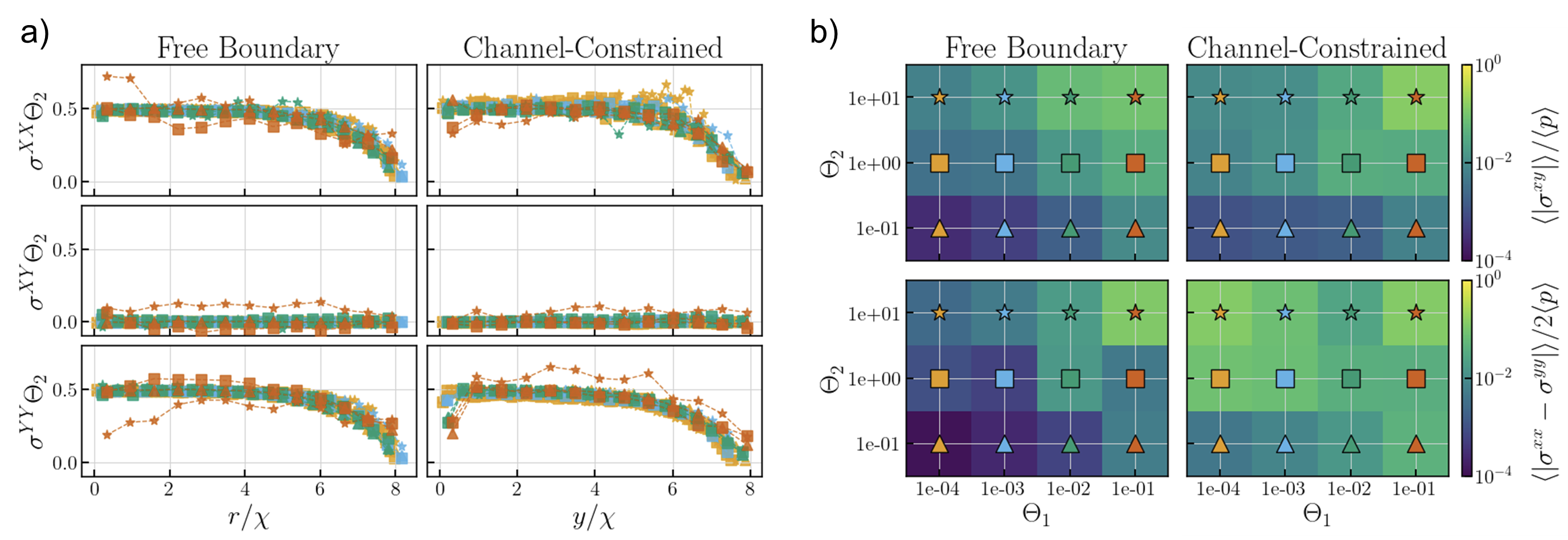}
\caption{
(a) Radial (left) and longitudinal (right) profiles of the stress tensor components $\sigma^{xx}$, $\sigma^{xy}$, and $\sigma^{yy}$, evaluated at a fixed colony size $R = 8\chi$ (free boundary, left column) and $L = 8\chi$ (channel-constrained, right column). (b) Phase diagrams as a function of $\Theta_1$ and $\Theta_2$. The top row shows the normalized shear $\langle |\sigma^{xy}| \rangle / \langle p \rangle$, while the bottom row shows the normalized anisotropy $\langle |\sigma^{xx}-\sigma^{yy}| \rangle / 2\langle p \rangle$. The left column corresponds to free boundary conditions ($R = 8\chi$), and the right column to channel-constrained growth ($L = 8\chi$).
}
\label{offDiagonalStress}
\end{figure}

In our simulations, these components remain small compared to the diagonal ones, as shown in Fig.~\ref{offDiagonalStress}, indicating that the colony is predominantly isotropic at coarse-grained scales. Under this approximation, the stress tensor reduces to

\begin{equation}\label{Adeviatoric}
    \pmb{\sigma}(\vec{r},t) = -p(\vec{r},t)\pmb{I}_2, 
\end{equation}
where $p(\vec{r},t)$ is the pressure, and $\pmb{I}_2$ is the $2 \times 2$ identity matrix.

\noindent \textbf{Darcy-like relation.} — Substituting Eqs.~\eqref{Adeviatoric} and \eqref{A2Dbodyforce1} into the force balance Eq.~\eqref{A2DNewton}, we obtain

\begin{equation}
\label{DarcyUP}
\boxed{
\vec{u}(\vec{r},t) = -\frac{\pi a \rho_0}{\mu \rho(\vec{r},t)} \nabla p(\vec{r},t)
}
\end{equation}
Which corresponds to Eq. (6) in the main text.

\subsection{S2.3. Pressure from overlap interaction}

We now relate the coarse-grained overlap field to the pressure.

At the microscopic level, stresses arise from pairwise repulsive interactions. The coarse-grained stress tensor in a region $d\Omega$ is given by the overdamped virial expression

\begin{equation}
\pmb{\sigma}(\vec r,t) = \frac{1}{2 d\mathcal A} \sum_{i,j\in d\Omega} \vec r_{ij} \otimes \vec f_{ij},
\end{equation}
where $\vec f_{ij} = -K d_{ij} H\!\left(d_{ij}\right)\hat r_{ij}$ and $d_{ij}=a_i+a_j-r_{ij}$.

Using $\vec r_{ij} = (a_i + a_j - d_{ij})\hat r_{ij}$, we obtain

\begin{equation}
\pmb{\sigma}(\vec r,t)
= -\frac{K}{2 d\mathcal A}
\sum_{i\in d\Omega} \sum_{j\neq i}
(a_i + a_j - d_{ij}) d_{ij}
(\hat r_{ij}\otimes \hat r_{ij})
H\!\left(d_{ij}\right).
\end{equation}

Assuming weak polydispersity ($a_i \approx a_j \approx a$) and small overlaps ($d_{ij} \ll 2a$), we approximate

\begin{equation}\label{AsigmaAB}
     \sigma^{\alpha\beta}(\vec{r},t) \approx -\frac{K a}{d \mathcal{A}}
        \sum_{i\in d\Omega} \sum_{j \neq i}d_{ij} (r^\alpha_{ij}r^\beta_{ij})H\!\left(d_{ij}\right).
\end{equation}

\noindent \textbf{Pressure.} — The pressure is defined from the trace of the stress tensor,

\begin{equation}
p(\vec r,t) = -\frac{1}{2}\,\mathrm{Tr}\,\pmb{\sigma}.
\end{equation}
Using Eq.~\eqref{AsigmaAB} and the identity $\sum_{\alpha} \hat r_{ij}^\alpha \hat r_{ij}^\alpha = 1$, we obtain

\begin{equation}
p(\vec r,t)
= \frac{K a}{2 d\mathcal A}
\sum_{i\in d\Omega} \sum_{j\neq i}
d_{ij}
H\!\left(d_{ij}\right).
\end{equation}

Rewriting the double sum in terms of the total overlap per particle from \eqref{overlapperparticle}, yields

\begin{equation}
p(\vec r,t)
= \frac{K a}{2 d\mathcal A}
\sum_{i\in d\Omega} d_i.
\end{equation}
Finally, expressing the result in terms of coarse-grained fields we lead to Eq. (7) from the main text,

\begin{equation}
\boxed{
p(\vec r,t)
=\frac{K\,\rho(\vec r,t)}{2 a \pi \rho_0}\, d(\vec r,t)
}.
\label{PressureOverlap}
\end{equation}

\subsection{S2.4. Free expansion solution}

When the system evolves under open boundary conditions, the colony adopts a circular shape of radius $R(t)$. Exploiting this axial symmetry, we introduce polar coordinates $\mathbf{r}=(r,\theta)$ and assume $p(\mathbf{r},t)=p(r,t)$. The Laplacian then reduces to

\begin{equation}
\nabla^2 p(r,t) = \frac{1}{r} \frac{\partial}{\partial r}\left[r\frac{\partial p}{\partial r}\right].
\end{equation}
Substituting this expression into Eq.~(9) of the main text yields an inhomogeneous modified Bessel equation. Its general solution can be written as

\begin{equation}
p(r,t) = p_c + C_1(t)\, I_0\!\left(\frac{r}{\chi}\right) + C_2(t)\, K_0\!\left(\frac{r}{\chi}\right),
\label{sol_open}
\end{equation}
where $I_0(x)$, and $K_0(x)$ are the Modified Bessel functions of first and second kind respectively. Now, noticing that $\displaystyle{\lim_{x \to 0}K_0(x) = \infty}$ the field's divergence at $r=0$ its avoided taking $C_2(t)=0$. On the other hand, the pressure at the colony's edge must be null in absence of external media, so $p(r=R(t),t) = 0$. Considering these conditions the pressure field results:
\begin{eqnarray}\label{overlapField_open}
    p(r,t) = p_c\left( 
    1 - \dfrac{I_0(r/\chi)}{I_0 \left( R(t)/\chi \right)}
    \right).
\end{eqnarray}

Therefore, by using \eqref{DarcyUP} with $\rho=\rho_0$, the velocity field is:

\begin{eqnarray}\label{overlapField_open}
    \vec{u}(r,t) = u_0 \dfrac{I_0(r/\chi)}{I_0 \left( R(t)/\chi \right)} \hat{r},
\end{eqnarray}
where $u_0=2\nu \chi/a$.

\noindent \textbf{Front dynamics.} —
For the radially symmetric colony, the interface position $R(t)$ is
determined by the velocity at the colony edge, $u_r(r=R(t),t)$.
Using the velocity field obtained from the pressure solution, the front dynamics follows

\begin{equation}
\dot{R}(t)=
u_0
\frac{I_1(R(t)/\chi)}{I_0(R(t)/\chi)} .
\end{equation}
Although we could not find any admissible simple closed-form solution, its limiting behaviors can be obtained analytically. For small colonies $R \ll \chi$, using that $u_0/\chi=2\nu/a=2/t^*$ and the expansions
$I_0(x)\approx1$ and $I_1(x)\approx x/2$, we obtain

\begin{equation}
\dot{R}(t) \approx R(t)/t^*,
\end{equation}
which leads to exponential growth,

\begin{equation}
R(t) \sim \exp\left(t/t^*\right).
\end{equation}
In the opposite limit $R \gg \chi$, since the modified Bessel functions
satisfy $I_1(x)/I_0(x)\to1$, the front is given by

\begin{equation}
\dot{R}(t) = u_0,
\end{equation}
and therefore a linear expansion of the colony radius. These two limits correspond to the exponential and constant-velocity
growth regimes observed in the discrete simulations.

\subsection{S2.5. Channel-limited expansion}

We now consider a confined geometry where the colony expands along a straight channel of width $w$, with periodic boundary conditions in the transverse direction $\hat{x}$ and propagation along $\hat{y}$. Due to translational invariance along $x$, the pressure field depends only on the longitudinal coordinate, $p(\vec r,t)=p(y,t)$. In this case the Laplacian reduces to

\begin{eqnarray}
\label{lapla_channel}
\nabla^2 p(y,t)=\frac{\partial^2 p(y,t)}{\partial y^2}.
\end{eqnarray}
Substituting Eq.~(\ref{lapla_channel}) into Eq.~(9) of the main text yields the one–dimensional inhomogeneous Helmholtz equation, with general solution

\begin{eqnarray}
\label{sol_channel}
p(y,t)=p_c + C_1(t)e^{y/\chi}+C_2(t)e^{-y/\chi}.
\end{eqnarray}
To determine the coefficients we impose the boundary conditions for the expanding colony. First, in absence of external media, the closed end of the channel requires vanishing pressure gradient,
$\partial_y p(0,t)=0$. Second, the pressure must vanish at the colony front,
$p(y=L(t),t)=0$. Applying these conditions to Eq.~(\ref{sol_channel}) yields

\begin{eqnarray}
p(y,t)=p_c\left[
1-\frac{\cosh\left(y/\chi\right)}
{\cosh\left(L(t)/\chi\right)}
\right].
\end{eqnarray}
Using our Darcy's relation Eq.~(\ref{DarcyUP}) with $\rho=\rho_0$, the longitudinal velocity field follows as

\begin{eqnarray}
\vec u(y,t)=
u_0
\frac{\sinh\left(y/\chi\right)}
{\cosh\left(L(t)/\chi\right)}
\,\hat y .
\end{eqnarray}

\noindent \textbf{Front dynamics.} —
The evolution of the colony length follows from the kinematic boundary
condition (KBC), which states that the velocity of the interface equals
the normal velocity of the material at the front. For the planar interface
located at $y=L(t)$ this condition reads

\begin{equation}
\dot{L}(t) = u_y(y=L(t),t).
\end{equation}
Substituting the velocity field obtained from Darcy's law yields

\begin{equation}
\dot{L}(t) = u_0 \tanh\!\left(\frac{L(t)}{\chi}\right),
\end{equation}
which governs the time evolution of the colony length. This equation can
be integrated analytically by separating variables,
$\coth\!\left(L/\chi\right) dL = u_0\,dt$, leading to the explicit solution

\begin{equation}
L(t) =
\chi \, \mathrm{arcsinh}
\left[
\exp\left(\frac{u_0 t}{\chi}\right)
\sinh\left(\frac{L_0}{\chi}\right)
\right],
\label{eq:L(t)}
\end{equation}
where $L_0=L(0)$ is the initial colony length.

This solution naturally captures the two growth regimes observed in the
simulations. For small colonies $L \ll \chi$, expanding
$\tanh(L/\chi) \approx L/\chi$ gives

\begin{equation}
\dot{L}(t) \approx \frac{u_0}{\chi}L(t),
\end{equation}
which leads to exponential growth $L(t) \sim \exp\left(\frac{u_0}{\chi}t\right)$.

In contrast, for large colonies $L \gg \chi$ we have
$\tanh(L/\chi)\to1$, so the front propagates at constant velocity,
\begin{equation}
\dot{L}(t) \approx u_0 ,
\end{equation}
yielding the asymptotic linear expansion. 

Thus the screening length $\chi$ controls the crossover between an
initial exponential expansion and a late-time regime of constant
front velocity.

\section{S4. Nutrient-Mechanical Link and General stability analysis}

\subsection{S4.1 Unified field formulation.}
Both mechanical and nutrient-based descriptions can be expressed within a unified framework by introducing a scalar field $q_\eta(\mathbf r,t)$, where the switch parameter $\eta\in\{0,1\}$ switches between the mechanical ($\eta=0$) and the nutrient-based model ($\eta=1$). The field satisfies Eq.~(13) of the main text in the passive ($\Omega^+$) and active ($\Omega^-$) regions, which are coupled through continuity of the field and its normal flux at the interface,
\begin{align*}
q_\eta(\partial\Omega^-) &= q_\eta(\partial\Omega^+),\\
\kappa_\eta \nabla q_\eta(\partial\Omega^+)\!\cdot\!\hat n
&=
\nabla q_\eta(\partial\Omega^-)\!\cdot\!\hat n ,
\end{align*}
with contrast parameter $\kappa_\eta$, together with the outer boundary condition
\begin{align*}
q_\eta|_{\partial\Omega_{\text{out}}^+} &= 0,\\
\lim_{\substack{\mathcal{D}(\mathbf{r},\partial\Omega)\to\infty \\\mathbf{r}\in\Omega_-}}
\nabla q_\eta(\mathbf{r},t) &= \mathbf{0}.
\end{align*}
where $\mathcal{D}(\mathbf{r},\partial\Omega)$ is the distance between $\mathbf{r}$ and the boundary $\partial\Omega$

\subsection{S4.2 General linear stability analysis}

We analyze the stability of a planar interface in a channel located at $y=y_I$. The stationary solution of the field is

\begin{equation}
q_\eta^{(0)}(\mathbf{r},t) =
\begin{cases}
\dfrac{q_I}{\Delta}(L-y), & y_I<y\le L, \\
q_0+(q_I-q_0)e^\frac{y-y_I}{\chi}, &y<y_I,
\end{cases}
\end{equation}
where $\Delta=L-y_I$ and $q_I=q_\eta(y_I)=q_0\left( 1+\kappa_\eta\chi/\Delta \right)^{-1}$ from boundary continuity flux. To probe the stability of this configuration we introduce a small perturbation of the interface,

\begin{equation}
y_I \rightarrow \Gamma_I(x,t) = y_I + \epsilon \xi(t)e^{ikx}, \qquad \epsilon \ll 1,
\label{supeq:perturbation}
\end{equation}
with $k=2\pi/\lambda$. Expanding the field as $q_\eta=q_\eta^{(0)}+\epsilon q_\eta^{(1)}e^{ikx}$, the first–order correction reads

\begin{equation}
q_\eta^{(1)}(\mathbf{r},t) =
\begin{cases}
A\sinh(k(L-y)), & y_I<y\le L, \\
Ce^{\tilde{k}(y-y_I)}, &y<y_I,
\end{cases}
\end{equation}
where $\tilde{k}=\sqrt{k^2+\chi^{-2}}$. Applying continuity of the field at the interface gives

\begin{equation}
C=
A\sinh(k\Delta)+\xi(t)q_0\dfrac{\kappa_\eta-1}{\Delta+\kappa_\eta\chi},
\end{equation}
while continuity of the normal flux yields

\begin{equation}
A=\xi(t)q_0
\dfrac{\kappa_\eta+\tilde{k} \chi \left(1-\kappa_\eta \right)}
{\chi\left(\Delta+\kappa_\eta\chi\right)\left(\kappa_\eta k\cosh(k\Delta)+\tilde k \sinh(k\Delta)\right)}.
\end{equation}
The interface dynamics is determined by Darcy's law

\begin{equation}
\vec{u}=-k_p\nabla \tilde{p}_\eta,
\end{equation}
combined with the kinematic boundary condition. Linearizing around the perturbed interface yields

\begin{equation}
\partial_y\tilde{p}_\eta^{(1)}(y)\big|_{y=y_I}
 = -e^{ikx}\left[
 \xi(t)\,\partial_y^2\tilde{p}_\eta^{(0)}\big|_{y=y_I}
 + \dot{\xi}(t)k_p^{-1}
 \right],
\end{equation}
which leads to the dispersion relation

\begin{equation}
\dfrac{\dot{\xi}}{\xi}=-k_p\left(
\partial_y\tilde{p}_\eta^{(1)}e^{-ikx}\xi^{-1}+\partial^2_y\tilde{p}_\eta^{(0)}
 \right).
\label{supeq:linearizeddynamics}
\end{equation}
To close the problem, a relation between the scalar field $q$ and the effective pressure $\tilde p$ must be specified. We use the unified expression $\tilde{p}_\eta=q_\eta-\eta(\alpha n_0 - \phi)$, introduced in Eq.~(14) of the main text. Here, the field $\phi$ arises from expressing the pressure in terms of the nutrient field $n$ and satisfies $\nabla^2\phi=0$, together with the boundary condition $\tilde{p}_1=0$ at the interface $\partial \Omega^-$. These conditions determine $\phi$ uniquely. At zeroth order, $\phi^{(0)}=\alpha n_0 - q_I = \alpha n(y_I)$, while at first order $\phi^{(1)}(y)=e^{k(y-y_I)}\left( q_\eta^{(1)}(y_I) + \xi \partial_yq_\eta^{(0)}(y_I) \right)$. Substituting this into Eq.~\eqref{supeq:linearizeddynamics}, we obtain the dispersion relation reported in Eq.~(15) of the main text,

\begin{equation}
\sigma_\eta(\lambda') = -k_p \left(
\xi^{-1} \partial_y q_\eta^{(1)} + \partial_y^2 q_\eta^{(0)} - \dfrac{2\pi\eta}{\lambda}\left(
\xi^{-1} q_\eta^{(1)} + \partial_y q_\eta^{(0)}
\right)
\right)\bigg|_{y_I}.
\end{equation}
The switch parameter $\eta$ allows to write two physical scenarios within the same framework, $\eta=0$ recovers the purely mechanical model, while $\eta=1$ corresponds to the nutrient-driven case. 

Differences between mechanical and nutrient-based models emerge beyond this regime, where the coupling term proportional to $\sigma$ modifies the feedback between interface perturbations and growth. This leads to quantitative differences in the dispersion relation and therefore in the selection of unstable modes. 

\begin{figure}[h!]
\centering
\includegraphics[width=0.7\textwidth]{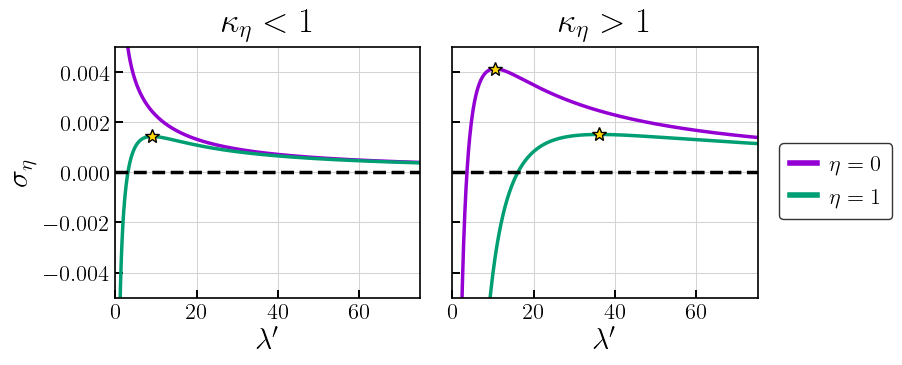}
\caption{Dispersion relation both nutrient ($\eta=1$) and mechanical ($\eta=0$) models in two different regimes  $\kappa_\eta=0.5$ (left) and $\kappa_\eta=2$ (right). Stars indicates the most unstable modes for $\lambda'>0$.}
\label{dispersion}
\end{figure}

The resulting dispersion relations are shown in Fig.~\ref{dispersion}. While both models share a similar structure for $\kappa_\eta>1$, the additional $\eta$-coupled term modifies the relation between interfacial perturbations and growth, leading to quantitative differences in the unstable spectrum and in the selection of the fastest-growing mode.

\section{Growth rate acceleration}

We derive the correction to the total growth rate induced by a small interfacial perturbation. The total area occupied by the colony is

\begin{equation}
A(t) = \int_{\Omega^-} d\mathbf{s},
\end{equation}
and its time evolution is geometrically governed by the divergence of the velocity field,

\begin{equation}
\frac{dA}{dt}
=
\int_{\Omega^-} \nabla \cdot \mathbf{v}\, d\mathbf{s}.
\end{equation}
Inside the colony, the production of cells is given by the sink

\begin{equation}
\nabla \cdot \mathbf{v} = -\zeta\nabla^2 p = \frac{\zeta}{\chi^2}(p_c-p)=\frac{2\nu}{a}(1-\dfrac{p}{p_c}),
\end{equation}
By using the intensive growth rate $g(t)$ defined in Eq. (11) from the main text, we write

\begin{equation}
g(t)
=
\frac{2\nu}{a\chi}\dfrac{1}{w}
\int_{\Omega^-} (1-\dfrac{p}{p_c})\, d\mathbf{s}.
\end{equation}

\noindent \textbf{Perturbed interface.} — We introduce a small perturbation of the interface as in \eqref{supeq:perturbation} but writing explicitly the real part so $\Gamma_I(x,t)=y_I + \epsilon \xi(t) \cos(kx)$. Therefore in a perturbed front in a channel geometry the colony domain is given by

\begin{equation}
\Omega^- = \{(x,y): 0 \le x \le w,\; 0 \le y \le \Gamma_I(x,t)\},
\end{equation}
so that

\begin{equation}
g(t)
=
\frac{2\nu}{a\chi}\dfrac{1}{w}
\int_0^w \int_0^{\Gamma_I(x,t)} (1 - \dfrac{p(x,y,t)}{p_c})\, dy\, dx.
\end{equation}

Now, we define the scalar field $f(x,y,t) \equiv 1 - p(x,y,t)/p_c$ and expand it as $f(x,y,t) = f^{(0)}(y) + \epsilon\, \xi(t)\, \hat f^{(1)}(y)\,  \cos(kx) $, where $\hat f^{(1)}(y)$ captures the spatial structure of the first-order correction, while $\xi(t)$ contains its time-dependent amplitude. This decomposition makes explicit that the field responds linearly to the interfacial perturbation.

The integral over the perturbed domain can be expanded as

\begin{equation}
\int_0^{y_I + \epsilon \xi(t) \cos(kx)} f(x,y,t)\, dy
=
\int_0^{y_I} f(x,y,t)\, dy
+ \epsilon \xi(t) \cos(kx) f(y_I)
+ \frac{\epsilon^2}{2} (\xi(t) \cos(kx))^2\partial_y f(y_I)
+ \mathcal{O}(\epsilon^3),
\end{equation}
Therefore, we expand the intensive growth rate as $g(t) = g^{(0)}(t)+\epsilon g^{(1)}(t) + \epsilon^2 g^{(2)}(t)$. At zeroth order we get the growth rate of a flat interface,

\begin{equation}
g^{(0)}(t)
=
\frac{2\nu}{a\chi}
\int_0^{y_I} f^{(0)}(y) \, dy=g_{\mathrm{flat}}.
\end{equation}
Defining the spatial average along the interface as  $\langle\cdot \rangle =\frac{1}{w}\int_0^w(\cdot)\,dx$, we found that at first order, the contribution reads:

\begin{equation}
g^{(1)}(t)=
\frac{2\nu}{a\chi}\xi(t) \langle \cos(kx) \rangle 
\left[
 \int_0^{y_I} \hat{f}^{(1)}(y)\, dy
+
\, f^{(0)}(y_I)
\right],
\end{equation}
which vanishes upon spatial averaging $\langle \cos(kx) \rangle = 0$.
At second order, non-vanishing contributions arise from quadratic terms in the interface perturbation. The dominant terms read

\begin{equation}
g^{(2)}(t)
=
\frac{2\nu}{a\chi}  \langle \cos(kx)^2  \rangle \xi(t)^2
\left[
\hat{f}^{(1)}(y_I)+\dfrac{1}{2} \partial_yf^{(0)}(y_I)
\right].
\end{equation}
Using $\langle \cos^2(kx) \rangle = 1/2$, only the squared amplitude of the perturbation contributes to the mean growth.
Collecting terms, for a generic perturbation dominated by the fastest-growing mode $k^*$, with growth rate $\sigma_\text{max}=\sigma(k^*)$, the intensive growth rate reads

\begin{equation}
g(t)
=
g_\text{flat}
+
\frac{\nu}{a\chi} B(k^*) \xi_0^2 e^{2\sigma_\text{max}t},
\end{equation}
where $B(k) =
\hat{f}^{(1)}(y_I)+\dfrac{1}{2} \partial_yf^{(0)}(y_I).$

Now, stable systems only have the zeroth order planar contribution, then $g_\text{stable}=g^{(0)}=g_\text{flat}$. Therefore, we compare the differences between growth rates of unstable and stable colonies defining $\Delta g \equiv g_{\mathrm{unstable}} - g_{\mathrm{stable}}$, then in the linear regime

\begin{equation}
\Delta g = \frac{\nu}{a\chi} B(k^*) \xi_0^2 e^{2\sigma_\text{max}t},
\end{equation}
leading to Eq. (12) of the main text.